\documentclass[aps,prl,twocolumn,superscriptaddress,nofootinbib]{revtex4-2}
\usepackage{amsmath,amssymb,bm,graphicx,hyperref,booktabs}
\hypersetup{colorlinks=true,linkcolor=blue,citecolor=blue,urlcolor=blue}

\begin{document}

\title{Hamiltonian Stabilization of the Two-Photon Dicke Model}
\author{Pradip Laha}
\email{plaha@uni-mainz.de}
\affiliation{Institute of Physics, Johannes Gutenberg-Universit\"at Mainz, Staudingerweg 7, 55128 Mainz, Germany}
\author{Kingshuk Adhikary}
\email{kadhikary@kaist.ac.kr}
\affiliation{Department of Physics, KAIST, Daejeon 34141, Republic of Korea}
\author{Subhashish Banerjee}
\email{subhashish@iitj.ac.in}
\affiliation{Department of Physics, Indian Institute of Technology Jodhpur, Rajasthan 342030, India}
\author{Peter van Loock}
\email{loock@uni-mainz.de}
\affiliation{Institute of Physics, Johannes Gutenberg-Universit\"at Mainz, Staudingerweg 7, 55128 Mainz, Germany}

\begin{abstract} 
Collective two-photon light--matter interactions enable pair amplification but can destroy spectral confinement, preventing a stable thermodynamic bright phase. We show that a positive Kac-scaled number-Kerr interaction renders the two-photon Dicke Hamiltonian of \(N\) emitters lower-bounded 
at all couplings while preserving its generalized \(\mathbb Z_4\) symmetry. We determine the exact zero-temperature thermodynamic energy density with \(O(N^{-1})\) finite-size control and obtain the global phase diagram.
A bright minimum is born at a saddle-node and exchanges global stability with the normal minimum at coexistence, producing a discontinuous photon-density jump. Symmetry-resolved exact diagonalization through \(N=128\) supports the predicted density jump and an exponentially closing same-sector normal--bright tunneling gap. Spectral collapse thus becomes the singular zero-Kerr point of a first-order phase diagram, identifying a Hamiltonian route to stabilizing pair amplification.

\end{abstract}
 
\maketitle

\textit{Introduction.}---
Collective light--matter interactions can generate macroscopic quantum phases from microscopic degrees of freedom.  The archetypal Dicke model realizes this mechanism through the collective coupling of a bosonic mode to \(N\) two-level emitters~\cite{Dicke1954,HeppLieb1973,WangHioe1973,EmaryBrandes2003PRL,Brandes2005,Brandes2013,Kirton2019,RosesDallaTorre2020,LarsonMavrogordatos2021,Villasenor2024DickeReview}. Above a critical coupling, this model undergoes a superradiant transition with macroscopic cavity occupation and spontaneous symmetry breaking~\cite{EmaryBrandes2003PRE,Lambert2004,Vidal2006,Dimer2007,Nagy2010,Baumann2010,Baumann2011,Ritsch2013,Brennecke2013,Klinder2015,Tiwari_prsa_2023,HolzingerGenes2025,Holzinger_PRA_2026}. Such a thermodynamically well-defined ordered phase, however, presupposes that the Hamiltonian remains spectrally confined.

Quadratic light--matter couplings challenge this prerequisite. In the bare two-photon Dicke model, the interaction changes the cavity occupation in pairs~\cite{GerryTogeas1989,LoLiuNg1998,Garbe2017,Peng2017DarkLike,Chen2018,Wang2019ChaosTPD,Banerjee2022Squeezing,Ramirez2025Integrability} and inherits spectral-collapse behavior from two-photon Rabi models and their multiqubit extensions~\cite{EmaryBishop2002,Duan2016,Felicetti2015,Felicetti2018,Lo2020,Rico2020,Lo2021Multiqubit,Braak2026KPhoton}. At the collapse threshold, one field quadrature loses confinement, and above it the energy becomes unbounded from below. 
The apparent bright regime of the bare model disappears in the thermodynamic limit, leaving the normal branch up to collapse~\cite{Ramirez2026FiniteSize}. 
Stabilization is therefore logically prior to identifying an ordered phase.

Open-system studies make this requirement explicit. 
In the bare model, ordinary one-photon loss supplies linear damping but no nonlinear field damping to saturate pair amplification.
In contrast, engineered two-photon loss supplies the missing nonlinear feedback, producing a stable dissipative transition~\cite{Shah2025}.  Related two-photon Dicke variants exhibit bistability, nonlinear attractors, and chaotic dynamics~\cite{Garbe2020,LiFazioChesi2022,LiChesi2024,Mylnikov2022}. 
Engineered nonlinear loss can therefore stabilize the open dynamics while leaving the closed Hamiltonian unbounded.
Here, we ask whether symmetry-preserving Hamiltonian confinement can remove the instability and support a genuine collective bright phase at fixed emitter and cavity frequencies, 
without nonlinear reservoir engineering.

We present a symmetry-preserving Hamiltonian regularization of the two-photon Dicke model by adding a positive, Kac-scaled number-Kerr interaction.  Higher-order Hamiltonian stabilizers have addressed spectral instabilities in multiphoton qubit--oscillator models~\cite{Ayyash_PRA_2025}, while Stark couplings tune two-photon Dicke transitions~\cite{Zhai2025} and quadrature-selective quartic potentials stabilize two-photon Rabi models~\cite{Ying2025}.

The Kerr interaction confines the field amplitude without an additional phase preference, preserving the generalized \(\mathbb Z_4\) symmetry and contributing extensively at finite photon density.  The normal minimum remains locally stable, while a finite-density bright branch emerges at a saddle-node and becomes degenerate with it at a distinct coexistence coupling, producing a discontinuous photon-density jump. We prove lower boundedness at every finite \(N\) and coupling, and determine the exact thermodynamic ground-state energy density with \(O(N^{-1})\) finite-size control.  Symmetry-resolved exact diagonalization (ED) through \(N=128\) corroborates this phase structure and supports exponential closure of the same-sector normal--bright tunneling gap.  Circuit-QED schemes for positive Kerr and pair interactions offer a concrete route toward implementing this Hamiltonian stabilization mechanism~\cite{Ye2025Quarton,Ayyash2024}.

\textit{Kerr-confined two-photon Dicke Hamiltonian.}---
The bare two-photon Dicke Hamiltonian describing \(N\) identical two-level emitters, each with frequency \(\omega_0\), collectively coupled to a bosonic mode of frequency \(\omega_c\) reads
\begin{equation}
  H_{\rm bare} = \omega_c a^\dagger a+\omega_0J_z + \frac{\lambda}{N} \left(a^2+a^{\dagger2}\right)J_x ,
  \label{eq:bare}
\end{equation}
with \(\hbar=1\). Here, \(a\) (\(a^\dagger\)) annihilates (creates) a bosonic excitation, and \(J_\mu=\frac12\sum_{k=1}^{N}\sigma_\mu^{(k)}\), with \(\sigma_\mu^{(k)}\) the Pauli operator of emitter \(k\), describes the collective spin (\(\mu=x,y,z\)). The interaction couples collective spin flips to pairs of bosonic excitations with collective strength \(\lambda\).  Since the phase rotation \(a\to ia\) reverses the sign of \(a^2+a^{\dagger2}\), we take \(\lambda\geq0\) without loss of generality.  

Introducing the field quadratures \(Q=(a+a^\dagger)/\sqrt2\) and \(P=(a-a^\dagger)/(i\sqrt2)\), the variational spin choice \(J_x=-N/2\) gives \(
   H_{\rm field}=\frac{1}{2}(\omega_c+\lambda)P^2 + \frac{1}{2}(\omega_c-\lambda)Q^2 -\frac{\omega_c}{2}. 
\)
At \(\lambda=\omega_c\), the  restoring stiffness along \(Q\)
vanishes, producing spectral collapse. For \(\lambda>\omega_c\), it becomes negative, and \( H_{\rm bare}\) becomes unbounded from below~\cite{Felicetti2015}.

We regularize this instability at the Hamiltonian level by adding
a Kac-scaled positive number-Kerr interaction
\begin{equation}
  H = H_{\rm bare} + \frac{\chi}{2N}\hat n^2 ,
  \label{eq:Hmain}
\end{equation}
where \(\chi>0\) is the strength of the intensive Kerr nonlinearity and $\hat n = a^\dagger a$ is the photon-number operator. 

Combining the quadrature inequality  \(-(2\hat n+1)\leq a^2+a^{\dagger2}\leq2\hat n+1\) with spin operator bounds \(\|J_x\|\leq N/2\) and \(J_z\geq-N/2\) gives
\begin{equation}
H\geq
-\frac{N}{2\chi}(\lambda-\omega_c)_+^2
-\frac{\omega_0N}{2}-\frac{\lambda}{2},
\label{eq:main_operator_bound}
\end{equation}
where \(x_+=\max(x,0)\).  Thus, the Hamiltonian \(H\) is bounded from below for every finite \(N\) and coupling, without restricting the spin polarization [see Supplemental Material (SM), Sec.~\ref{sec:SM_operator_boundedness} for details].

The normally ordered Kerr form \(a^{\dagger2}a^2=\hat n^2-\hat n\) differs only by the finite-\(N\) cavity-frequency shift \(\delta\omega_c=-\chi/(2N)\), which can be absorbed into the  detuning.

Both nonlinear terms in \(H\) in Eq.~\eqref{eq:Hmain} are Kac-normalized. At finite photon density, \((a^2+a^{\dagger2})J_x\) and \(\hat n^2\) are \(O(N^2)\), so their \(1/N\) coefficients produce an extensive Hamiltonian.  
For a general Kerr term \(K_N\hat n^2/2\),
\(K_N\sim\chi/N\) is the marginal scaling that retains finite, nonzero competition with pair amplification at fixed photon density and fixed \(\omega_c,\omega_0,\lambda\).
Normal ordering, scaling trichotomy, and circuit-QED emulation are discussed in SM, Secs.~\ref{subsec:normal_order}, \ref{sec:SM_Kac_trichotomy}, and \ref{sec:SM_circuit_QED}, respectively.




\textit{Exact thermodynamic energy landscape.}---
We determine the zero-temperature thermodynamics of Eq.~\eqref{eq:Hmain} using bosonic coherent states (CSs) \(|\sqrt{N}\,\alpha\rangle\), with \(\alpha=\sqrt n\,e^{i\phi}\). Here \(n=|\alpha|^2\) is the photon density 
and \(\phi\) is the bosonic phase.
At fixed \(n\) and \(\phi\), minimizing the leading CS energy density over the collective-spin orientation gives the spin-minimized energy density
\begin{equation}
  e(n,\phi) = \omega_c n+\frac{\chi}{2}n^2 - \frac12 \sqrt{ \omega_0^2+ 4\lambda^2n^2\cos^2(2\phi) }.
  \label{eq:e2_phase}
\end{equation}
Minimization over \(\phi\) selects \(|\cos(2\phi)|=1\), yielding
\begin{equation}
  e(n) = \omega_c n+\frac{\chi}{2}n^2 - \frac12 R(n),
  \label{eq:e2main}
\end{equation}
where \(R(n)=\sqrt{\omega_0^2+4\lambda^2n^2}\) and \(e(0)=-\frac{\omega_0}{2}\).  
Any finite-density minimum therefore has four \(\mathbb Z_4\)-related configurations at \(\phi=0,\pi/2,\pi,3\pi/2\). The CS construction and the spin orientation minimization are derived in SM, Secs.~\ref{sec:SM_exact_symbols} and \ref{sec:SM_spin_phase_minimization}. 

To establish the \emph{exactness} of Eq.~\eqref{eq:e2main}, let \(E_0(N)\) denote the ground-state energy in the full spin--boson Hilbert space. Constructing upper and lower CS symbols for the bosonic and spin degrees of freedom~\cite{Lieb1973Classical,LiebSeiringerYngvason2005}, we obtain the finite-\(N\) energy-density bounds
\begin{equation}
  \min_{n\geq0}e_N^-(n) \leq \frac{E_0(N)}{N} \leq \min_{n\geq0}e_N^+(n),
  \label{eq:main_exact_symbol_sandwich}
\end{equation}
where
\begin{align}
  e_N^+(n) &= e(n)+\frac{\chi n}{2N}, \label{eq:main_exact_symbol_bound_plus} \\
  e_N^-(n) &= e(n) - \frac{R(n)+\omega_c+3\chi n/2}{N} + \frac{\chi}{2N^2}.
  \label{eq:main_exact_symbol_bound_minus}
\end{align}
The upper bound follows from a product CS trial state, whereas the lower bound holds over the full Hilbert space, including entangled and squeezed states. 
For \(N\ge1\), \(e(n)\) and \(e_N^\pm(n)\) have a common quadratic lower bound with positive leading coefficient \(\chi/2\) and uniformly bounded linear coefficients. Their minimizers therefore lie in an \(N\)-independent compact interval, on which the symbol corrections are uniformly \(O(N^{-1})\). The two bounds converge to the same minimum \(e_\infty=\min_{n\geq0}e(n)\), i.e.,
\begin{equation}
  \lim_{N\to\infty}\frac{E_0(N)}{N}=e_\infty,
  \,\,\,
  \frac{E_0(N)}{N}=e_\infty+O(N^{-1}).
  \label{eq:main_exact_thermodynamic_limit}
\end{equation}
SM, Secs.~\ref{sec:SM_operator_boundedness} and \ref{sec:SM_exact_thermodynamic_limit}, give the operator bound, the exact maximal-spin reduction, and the CS proof. The associated Husimi measures concentrate on the CS configurations attaining the global minimum.
Hence, the branch analysis below determines the exact zero-temperature phase diagram rather than a mean-field approximation.

\begin{figure*}[t]
\centering
\includegraphics[width=\textwidth]{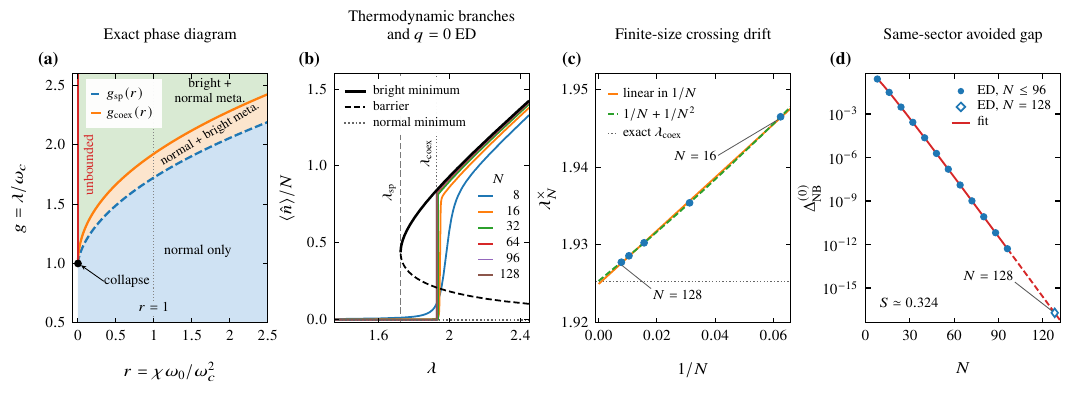}
\vspace{-5ex}
\caption{\textbf{Global phase diagram and finite-size signatures of the first-order normal--bright transition.}
(a) Exact zero-temperature phase diagram in the dimensionless \((r,g)\) plane, with \(r=\chi\omega_0/\omega_c^2\) and \(g=\lambda/\omega_c\). The blue dashed spinodal curve \(g_{\rm sp}(r)\) marks the saddle-node creation of a barrier and a bright local minimum, while the orange solid coexistence curve \(g_{\rm coex}(r)\) marks equality of the normal and bright energies.  For \(r>0\), the three regions contain, in order of increasing \(g\), only the normal minimum, a normal ground state with a metastable bright minimum, and a bright ground state with a metastable normal boundary minimum. The bare spectral-collapse point is at \((r,g)=(0,1)\). The red segment at \(r=0,\ g>1\) denotes a bare Hamiltonian unbounded from below.
Panels (b)--(d) use \(\omega_c=\omega_0=\chi=1\), corresponding to the \(r=1\) cut indicated in panel (a).
(b) Normal, barrier, and bright thermodynamic branches together with the photon density \(\langle\hat n\rangle/N\) of the lowest \(q=0\) state for the indicated system sizes.  The finite-size crossover sharpens toward the exact jump \(0\to n_{\rm coex}\).
(c) Finite-size drift of the branch-crossing coupling \(\lambda_N^\times\), where the normal and bright branch energies coincide.  A fit including \(1/N\) and \(1/N^2\) corrections gives \(\lambda_\infty^\times\simeq1.925365\), within \(4.06\times10^{-5}\) of the exact coexistence coupling.
(d) Same-sector normal--bright tunneling gap \(\Delta_{\rm NB}^{(0)}=E_1^{(0)}-E_0^{(0)}\) in \(q=0\), evaluated at \(\lambda_N^\times\). A fit over \(16\leq N\leq96\) to \(A N^p e^{-SN}\) gives \(S\simeq0.324\). Its extrapolation predicts the independently calculated \(N=128\) gap, excluded from the fit, within \(2\%\) (see main text).
}
\label{fig:main}
\end{figure*}

\textit{First-order normal--bright transition.}---
For the bare model (\(\chi=0\)), \(e(n)=(\omega_c-|\lambda|)n+O(1)\) at large \(n\).  Confinement is lost at $|\lambda|=\omega_c$, and above this coupling, the energy is unbounded from below.  
For \(\chi>0\), the asymptotic behavior \(e(n)\sim\chi n^2/2\) restores large-density confinement at every coupling.
Unlike the quartic potential $A_4(a+a^\dagger)^4$ used to regularize two-photon Rabi models~\cite{Ying2025}, the Kerr term confines the radial amplitude without adding a new angular anisotropy, thereby preserving the generalized $\mathbb{Z}_4$ symmetry.  
SM Secs.~\ref{sec:SM_context} and \ref{sec:SM_one_photon} compare these regularization mechanisms and contrast the first-order normal--bright transition with the continuous one-photon Dicke transition.

The $n=0$ normal state remains a one-sided boundary minimum for all couplings. Near the normal boundary
\begin{equation}
  e(n) = - \frac{\omega_0}{2}+\omega_c n + \left(\frac{\chi}{2}-\frac{\lambda^2}{\omega_0}\right)n^2 + O(n^4).
\end{equation}
Thus, \(\partial_n e|_{n=0}\equiv e'(0)=\omega_c>0\), so the normal minimum cannot lose stability continuously, excluding a second-order normal--bright transition. 
Instead, two finite-density stationary branches meet at a saddle-node. Solving \(e'(n_{\rm sp})=e''(n_{\rm sp})=0\) and selecting the positive-density solution gives the unique physical spinodal point
\begin{align}
 n_{\rm sp} &= \frac{\sqrt{\omega_c^2+(4\chi\omega_c\omega_0)^{2/3}} -\omega_c}{2\chi}.
\label{eq:spinodal}
\end{align}
Correspondingly, \(\lambda_{\rm sp} = \omega_c\left(1+\frac{\chi n_{\rm sp}}{\omega_c}\right)^{3/2}\). The two positive roots \(n_\pm\) of \(e'(n)=0\) coincide at the spinodal, \(n_\pm=n_{\rm sp}\).  For \(\lambda>\lambda_{\rm sp}\), they separate into a radial barrier at \(n_-<n_{\rm sp}\) and a bright local minimum at \(n_+>n_{\rm sp}\), while the normal boundary minimum at \(n=0\) persists.

The bright minimum is metastable for \(\lambda_{\rm sp}<\lambda<\lambda_{\rm coex}\), becomes degenerate with the normal minimum at coexistence, \(e(n_{\rm coex})=e(0)\), and is globally stable above~it.
Combining \(e(n)=e(0)\) with \(e'(n)=0\) for \(n>0\) gives
\begin{align}
\chi^2n^3+2\chi\omega_c n^2-2\omega_c\omega_0 =0.
\label{eq:coexistence}
\end{align}
Among the cubic's algebraic roots, the unique positive root defines \(n_{\rm coex}=n_+(\lambda_{\rm coex})\) and the positive square root of 
\(\lambda^2 = (\omega_c+\chi n)(2\omega_c+\chi n)^2/(4\omega_c)\)
fixes \(\lambda_{\rm coex}\). This physical solution gives the thermodynamic density jump \(0\to n_{\rm coex}\). At \(\lambda_{\rm coex}\), the lower positive stationary root \(n_b=n_-(\lambda_{\rm coex})\) forms the barrier separating the degenerate normal and bright minima. Thus, \(e(n_b;\lambda_{\rm coex})>e(0;\lambda_{\rm coex})=e(n_{\rm coex};\lambda_{\rm coex})\). 
For \(\chi>0\), the spinodal and coexistence conditions imply \(\omega_c<\lambda_{\rm sp}<\lambda_{\rm coex}\) [see SM, Eq.~\eqref{eq:SM_scale_ordering}]. Thus, the confined bright branch emerges beyond the bare collapse threshold.



These minima identify the thermodynamic ground-state configurations. Below \(\lambda_{\rm coex}\), the normal configuration \(n=0\) minimizes the energy, with the spin pointing along \(-z\) and \(\langle J_z\rangle/N\to-1/2\). Above coexistence, the four symmetry-related degenerate bright configurations \(\alpha=\pm\sqrt{n_+},\,\pm i\sqrt{n_+}\) have the lowest energy. They differ in phase but share the photon density \(n_+=|\alpha|^2\), each with its corresponding spin tilted away from \(-z\).  A symmetry-preserving ground-state sequence therefore has four equally weighted peaks in its limiting phase-space distribution.  An arbitrarily weak field favoring one bright configuration selects a symmetry-broken phase when removed after taking \(N\to\infty\).

Figure~\ref{fig:main}(a) summarizes the exact phase diagram in the dimensionless variables \(r=\chi\omega_0/\omega_c^2\) and \(g=\lambda/\omega_c\).  
Boundary formulas and their asymptotic limits are given in SM, Sec.~\ref{sec:SM_branch_selection}.
Panels~(b)--(d) of Fig.~\ref{fig:main} illustrate the \(r=1\) cut for \(\omega_c=\omega_0=\chi=1\), where \(g=\lambda\). The spinodal and coexistence occur at \((n_{\rm sp},\lambda_{\rm sp})\simeq(0.438,1.725)\) and \((n_{\rm coex},\lambda_{\rm coex})\simeq(0.839,1.925)\), respectively.  At coexistence, the barrier at \(n_b\simeq0.210\) has height \(\Delta e_b=e(n_b;\lambda_{\rm coex}) - e(0;\lambda_{\rm coex})\simeq0.0890\) per emitter. The corresponding CS energy barrier scales as \(N\Delta e_b+O(1)\).

At coexistence, the thermodynamic ground-state energy density \(e_\infty\) is continuous but has a kink, as \(\partial_\lambda e_\infty\) jumps from zero to \(-2\lambda_{\rm coex}n_{\rm coex}^2/R(n_{\rm coex})<0\). At fixed \(\omega_c,\omega_0>0\), the weak-Kerr limit \(\chi\to0^+\) gives \(g_{\rm sp},g_{\rm coex}\to1^+\), while \(n_{\rm sp}\) and \(n_{\rm coex}\) diverge.  The confined bright phase therefore terminates singularly at bare spectral collapse. On the \(r=0\) axis, \(g>1\) corresponds to an unbounded Hamiltonian, not a finite-density bright phase. Existence, uniqueness, and global branch selection are established in SM, Sec.~\ref{sec:SM_branch_selection}.

\textit{Finite-size scaling.}---
We now examine the approach to the exact thermodynamic limit through three complementary finite-size signatures: a sharpening photon-density crossover, convergence of the branch-crossing coupling, and closure of the normal--bright tunneling gap.

The exact maximal-spin reduction restricts the ground-state calculation to \(j=N/2\), where the generalized \(\mathbb Z_4\) symmetry is generated by \(\mathcal U_4= \exp\{\frac{i\pi}{2}[\hat n+2(J_z+j)]\}\). Its eigenvalues \(i^q\), \(q=0,1,2,3\), label four independent symmetry sectors.  In the basis \(|n_a,m\rangle\), with photon number \(n_a\) and \(J_z\) eigenvalue \(m\), \(q=[n_a+2(m+j)]\bmod4\). The normal branch evolves from the uncoupled vacuum with all spins down (\(n_a=0,m=-j\), hence \(q=0\)). Since \(q\) is conserved, it can hybridize only with bright states in the same sector; levels in different sectors cannot mix and may cross exactly. We therefore follow the lowest \(q=0\) eigenstate to isolate the normal--bright crossover.  Its cutoff-converged photon-density curves for \(N=8,16,32,64,96,128\), shown in Fig.~\ref{fig:main}(b), sharpen toward the thermodynamic jump.

To separate the normal and bright contributions near coexistence, let \(|\psi_0^{(0)}\rangle\) and \(|\psi_1^{(0)}\rangle\) denote the two lowest exact \(q=0\) eigenstates. Diagonalizing \(\hat n\) projected onto this doublet defines orthogonal low-density normal and high-density bright states, \(|{\rm N}\rangle\) and \(|{\rm B}\rangle\). Their Hamiltonian expectation values define the branch energies \(E_{\rm N}^{(N)}(\lambda)\) and \(E_{\rm B}^{(N)}(\lambda)\). 
The finite-size branch-crossing coupling occurs when \(E_{\rm N}^{(N)}(\lambda_N^\times) = E_{\rm B}^{(N)}(\lambda_N^\times)\).
These branch energies cross, but the exact eigenvalues remain separated there by \(\Delta_{\rm NB}^{(0)}=E_1^{(0)}-E_0^{(0)}\). At the crossing, the branch-density contrast is \(\Delta n_N = [\langle{\rm B}|\hat n|{\rm B}\rangle- \langle{\rm N}|\hat n|{\rm N}\rangle]/N\), which approaches \(n_{\rm coex}\). For \(N=128\), \(\Delta n_{128}\simeq0.833577\), within \(0.68\%\) of the thermodynamic jump. Checks of the two-state construction and the inter-sector splittings are given in SM, Secs.~\ref{sec:SM_symmetry_gap} and \ref{sec:SM_intersector_splittings}.

The branch-crossing coupling \(\lambda_N^\times\) approaches \(\lambda_{\rm coex}\) as \(N\) increases, with its finite-size shift arising from unequal quantum corrections to the normal and bright branch energies. Let \(e_{\rm N}=e(0)\) and \(e_{\rm B}(\lambda)=e[n_+(\lambda);\lambda]\). Their slope difference at coexistence is
\begin{equation}
  \ell_\lambda = -\left.\frac{d(e_{\rm B}-e_{\rm N})}{d\lambda} \right|_{\lambda_{\rm coex}}\!\!\!\! = \frac{(\omega_c+\chi n_{\rm coex})n_{\rm coex}} {\lambda_{\rm coex}}>0.
\end{equation}
Write \(E_\beta^{(N)}(\lambda) = Ne_\beta(\lambda)+\gamma_\beta(\lambda)+O(N^{-1})\), with \(\beta\in\{{\rm N},{\rm B}\}\). Here \(\gamma_{\rm N}\) and \(\gamma_{\rm B}\) are the leading \(O(1)\) quantum corrections to the total normal and bright branch energies, obtained from Gaussian fluctuations about their respective minima.
Balancing the branch energies gives
\begin{equation}
\lambda_N^\times = \lambda_{\rm coex}+\frac{c_1}{N}+O(N^{-2}),
\label{eq:finite_size_drift}
\end{equation}
with \(c_1 = [\gamma_{\rm B}(\lambda_{\rm coex}) -\gamma_{\rm N}(\lambda_{\rm coex})]/{\ell_\lambda}\). Thus an \(O(1)\) branch-energy mismatch produces an \(O(N^{-1})\) crossing shift, with \(\lim_{N\to\infty}\lambda_N^\times=\lambda_{\rm coex}\).

Figure~\ref{fig:main}(c) shows this convergence. For \(\omega_c=\omega_0=\chi=1\),  \(\ell_\lambda\simeq0.801781\). Fitting the five plotted crossings with a free intercept and both \(1/N\) and \(1/N^2\) corrections gives \(\lambda_\infty^\times\simeq1.925365\), within \(4.1\times10^{-5}\) of the exact coexistence coupling. For the number-squared Kerr interaction, \(\gamma_{\rm N}=0\), and the coupled Gaussian correction to the bright-well energy predicts \(c_1\simeq0.311401\). Fixing the exact intercept and fitting the ED data for \(N\geq40\) gives \(c_1^{\rm ED}\simeq0.311062\), in agreement within \(0.11\%\).  At \(N=128\), the crossing occurs at \(\lambda_{128}^\times\simeq1.927778\).

Figure~\ref{fig:main}(d) shows how tunneling between the normal and bright configurations is suppressed as \(N\) increases. At \(\lambda_N^\times\), the gap is \(\Delta_{\rm NB}^{(0)} = 2|\langle{\rm N}|H|{\rm B}\rangle|\). It directly measures the quantum mixing between the two branch states.  As this mixing vanishes, the finite-size crossover sharpens toward the thermodynamic density jump. Arbitrary-precision diagonalization with photon-cutoff checks resolves
the exponentially small gaps. Fitting the gaps over \(16\leq N\leq96\) to \(A N^p e^{-SN}\) gives \(S\simeq0.324\). The fit predicts the independently calculated \(\Delta_{\rm NB}^{(0)}(128)\simeq1.94\times10^{-17}\), excluded from the fit, within \(2\%\), supporting exponential tunneling suppression over the accessible sizes. Numerical convergence checks and fit details are given in SM, Sec.~\ref{sec:SM_ED_scaling}.

The small tunneling gap does not imply that either well loses local stability at coexistence.  Local softening instead occurs at the spinodal, where the bright minimum merges with its neighboring barrier. 
A large-\(N\) Gaussian expansion about the bright minimum gives \(\varepsilon_-\propto(\lambda-\lambda_{\rm sp})^{1/4}\) as \(\lambda\downarrow\lambda_{\rm sp}\), whereas both bright-well modes remain finite at coexistence. 
This local softening is therefore distinct from the interwell tunneling scale in Fig.~\ref{fig:main}(d) [see SM, Sec.~\ref{sec:SM_gaussian_fluctuations} for details].

\textit{Passive one-photon-loss dynamics.}---
Hamiltonian confinement does not by itself guarantee a dissipative bright attractor. We retain the positive Kerr term and include local Markovian cavity loss, independent spin relaxation, and pure dephasing, without incoherent pumping, nonlinear reservoirs, or feedback. In the resulting self-consistent mean-field dynamics, the normal fixed point remains locally attracting at every coupling for positive cavity loss and spin relaxation.

For \(\omega_c=\omega_0=\chi=1\), nonzero stationary photon densities appear in pairs above a decay-dependent coupling threshold. The lower-density branch is analytically unstable, whereas the upper branch can coexist as an attractor with the normal state. Equal longitudinal and transverse spin-decay rates exclude nonzero fixed points at Hamiltonian coexistence, but this exclusion does not extend to all relaxation--dephasing ratios. Moreover, the numerical stability analysis shows that increasing pure dephasing at fixed relaxation can destabilize the upper branch through an oscillatory mode before the two stationary branches merge and disappear. Bright-state existence and attraction are therefore distinct, and neither follows from Hamiltonian confinement alone. SM, Sec.~\ref{sec:SM_passive} derives the mean-field equations from the Lindblad model, establishes analytical existence bounds and lower-branch instability, and presents the full Jacobian analysis and decay-dependent stability diagrams. These conclusions concern local mean-field dynamics, not the exact finite-\(N\) quantum steady state.

\textit{Discussion.}---
A positive, Kac-scaled number-Kerr interaction converts the collapse-prone two-photon Dicke Hamiltonian into a confined collective model with an exact first-order normal--bright transition. The approach preserves the \(\mathbb Z_4\) symmetry, yields an exact collective thermodynamic description, and distinguishes local spinodal softening from normal--bright tunneling. At fixed coupling above the bare collapse threshold, the bright density diverges as \(\chi\to0^+\), so this phase is not a finite-density phase of the bare model. Kerr-induced Hamiltonian confinement thus complements the established dissipative stabilization through two-photon loss~\cite{Shah2025}.

The narrow normal--bright avoided crossing enhances ground-state sensitivity to coupling, but a small gap alone does not establish a sensing advantage without accounting for preparation time and calibration.
Single-spin multiphoton interactions support protocols for NOON and \(W\) states~\cite{Laha_MPJC_PRR_2024}, excitation transfer~\cite{Laha_MPJC_PRA_2025}, and binomial-code preparation~\cite{Laha_Binomial_PRR_2026}. Related Kerr-cat proposals and experiments combine Hamiltonian protection with frequency-selective dissipation to suppress leakage-induced errors~\cite{Putterman_PRL_2022,Ding_nat_com_2025}. Here, confinement requires no engineered reservoir. The phase boundaries, symmetry selection rules, and excitation gaps provide benchmarks for extending state preparation and control to many emitters. Protecting encoded quantum information against noise remains a separate requirement.

A driven circuit-QED emulator offers a route to finite-\(N\) spectroscopy and coherent transfer between the normal and bright branches, with parameter matching and effective-model corrections detailed in SM, Sec.~\ref{sec:SM_circuit_QED}. These tests probe the Hamiltonian branch structure rather than passive steady-state order. Extending the phase analysis to \(k\)-photon Dicke Hamiltonians (\(k>2\)) requires confinement that controls the interaction's \(n^{k/2}\) growth at large photon density. Number-Kerr confinement is marginal at \(k=4\) and insufficient for \(k>4\).

\textit{Acknowledgments.--} PL and PvL acknowledge funding from the BMFTR in Germany for support via PhotonQ, QR.N, QuKuK, QuaPhySI, and from the Deutsche Forschungsgemeinschaft  (DFG, German Research Foundation) – Project-ID 429529648 – TRR 306 QuCoLiMa (“Quantum Cooperativity of Light and Matter”).. KA was supported by KAIST through the Strategic Research Program (Project No. N10260106) and the C2 Project, as well as by the National Research Foundation of Korea (NRF) under Grant No. RS-2026-25483912.

 \textit{Disclosure of AI use.--} 
 P.L. conceived the central idea of this work almost two years ago, without AI assistance. Most of the key results had been obtained before the authors used ChatGPT-5.5 to examine whether the corresponding derivations could be reproduced. The AI model was also used to assist with refining mathematical arguments, writing codes, preparing figures, and editing the manuscript. The authors have verified all scientific results and take full responsibility for the content of the manuscript, including all AI-assisted contributions.

\bibliography{prl_refs}

\clearpage
\onecolumngrid
\setcounter{secnumdepth}{2}

\setcounter{section}{0}
\setcounter{subsection}{0}
\setcounter{secnumdepth}{2}
\setcounter{equation}{0}
\setcounter{figure}{0}
\setcounter{table}{0}
\renewcommand{\thesection}{S\arabic{section}}
\renewcommand{\thesubsection}{S\arabic{section}.\arabic{subsection}}
\makeatletter
\renewcommand{\p@subsection}{}
\makeatother
\renewcommand{\theequation}{S\arabic{equation}}
\renewcommand{\thefigure}{S\arabic{figure}}
\renewcommand{\thetable}{S\arabic{table}}
\renewcommand{\theHsection}{S.\arabic{section}}
\renewcommand{\theHsubsection}{S.\arabic{section}.\arabic{subsection}}
\renewcommand{\theHequation}{S.\arabic{equation}}
\renewcommand{\theHfigure}{S.\arabic{figure}}
\renewcommand{\theHtable}{S.\arabic{table}}

\begin{center}
{\large\bf Supplemental Material for ``Hamiltonian Stabilization of the Two-Photon Dicke Model''}\par\vspace{0.75em}
Pradip Laha, Kingshuk Adhikary, Subhashish Banerjee, and Peter van Loock \\
{\it Institute of Physics, Johannes Gutenberg-Universit\"at Mainz, Staudingerweg 7, 55128 Mainz, Germany}\\
{\it Department of Physics, KAIST, Daejeon 34141, Republic of Korea}\\
{\it Department of Physics, Indian Institute of Technology Jodhpur, Rajasthan 342030, India}
\end{center}

\vskip2ex

This Supplemental Material (SM) provides the derivations, numerical methods, and implementation details supporting the main text. Section~\ref{sec:SM_context} relates Kerr confinement to the bare model and dissipative, Stark, and quadrature-selective regularizations. Section~\ref{sec:SM_one_photon} contrasts the ordinary one-photon Dicke model with collective pair amplification. Sections~\ref{sec:SM_operator_boundedness}--\ref{sec:SM_branch_selection} establish finite-\(N\) confinement, the exact thermodynamic limit, marginal Kerr scaling, and the global phase diagram. Sections~\ref{sec:SM_ED_scaling} and~\ref{sec:SM_gaussian_fluctuations} present symmetry-resolved finite-size scaling, local Gaussian excitations and quantum correlations, and localized bright-state validation. Section~\ref{sec:SM_passive} derives the dissipative mean-field equations and distinguishes bright fixed-point existence from local stability under cavity loss and spin relaxation and dephasing. Finally, Section~\ref{sec:SM_circuit_QED} presents the proposed circuit-QED implementation and its Floquet parameter mapping.

\vskip5ex
\twocolumngrid

\section{Relation to past work and scope}
\label{sec:SM_context}

Background on the Dicke transition and its finite-size, entanglement, dynamical, and experimental signatures is given in Refs.~\cite{Dicke1954,HeppLieb1973,WangHioe1973,EmaryBrandes2003PRL,Brandes2005,Brandes2013,Kirton2019,RosesDallaTorre2020,LarsonMavrogordatos2021,Villasenor2024DickeReview,EmaryBrandes2003PRE,Lambert2004,Vidal2006,Dimer2007,Nagy2010,Baumann2010,Baumann2011,Ritsch2013,Brennecke2013,Klinder2015,Tiwari_prsa_2023}. Two-photon spectral collapse and related models are discussed in Refs.~\cite{EmaryBishop2002,Duan2016,Felicetti2015,Felicetti2018,Garbe2017,Chen2018,Lo2020,Wang2019ChaosTPD,Rico2020,Lo2021Multiqubit,Braak2026KPhoton,Peng2017DarkLike,Banerjee2022Squeezing,Ramirez2025Integrability}, and pair-driven open-system dynamics in Refs.~\cite{Garbe2020,LiFazioChesi2022,LiChesi2024,Mylnikov2022}. Here we clarify the distinctions relevant to the present analysis.

Reference~\cite{Ramirez2026FiniteSize} treats the bare two-photon model \(\chi=0\), whereas our results concern the positive-Kerr Kac sequence.  With \(\lambda=2\gamma\), their collapse threshold \(\gamma=\omega_c/2\) coincides with \(\lambda=\omega_c\). The conclusions are thus compatible. Section~\ref{sec:SM_branch_selection} describes the singular approach of the confined bright branch to bare collapse as \(\chi\to0^+\).

Hamiltonian confinement and reservoir stabilization address different problems.  The two-photon loss channel \(\kappa_2\mathcal D[a^2]\) can stabilize an open-system bright attractor~\cite{Shah2025}, whereas positive Kerr makes the closed Hamiltonian bounded from below.  We treat ordinary one-photon loss separately, testing passive fixed-point existence and attraction in Sec.~\ref{sec:SM_passive}, using standard open-system and driven-cavity methods~\cite{Keeling2010,Bhaseen2012,DallaTorre2013,Hwang2015,KirtonKeeling2017,Carmichael1999,GardinerZoller2004,BreuerPetruccione2002,Banerjee_open_system,Daley2014,Shammah2018}.

 Matter–matter interactions provide a distinct route to modifying superradiant thresholds and transition order in conventional one-photon Dicke models \cite{Mendonca_PRL_2025,jiao_2026}.
 Those studies do not address spectral collapse of the pair-coupled Hamiltonian considered here.

The form of the regularizer also matters.  Stark couplings modify the emitter splitting and transition onset in two-photon Dicke--Stark models~\cite{Zhai2025}, while a Cartesian quartic potential \(A_4(a+a^\dagger)^4\) stabilizes two-photon Rabi models by selecting a field quadrature~\cite{Ying2025}. The reduced scalar potential in Ref.~\cite{Ying2025}, Eq.~(9), maps to Eq.~\eqref{eq:e2main} here under \(x^2=2n\), \(\Omega=\omega_0\), \(\omega=\omega_c\), \(A_4=\chi/32\), and \(g_2=\lambda/4\). This correspondence does not equate the quantum models or their limiting procedures.  The Kerr term confines the amplitude without additional angular anisotropy and preserves the generalized \(\mathbb Z_4\) symmetry.

Higher-order Hamiltonian stabilizers also remove unbounded-below spectral instabilities while preserving the relevant low-photon sector in multiphoton qubit--oscillator models~\cite{Ayyash_PRA_2025}. In the one-photon Dicke model, negative Kerr lowers the superradiant threshold but leaves the closed system unstable, with cavity dissipation providing stabilization~\cite{schmid_2026}. These mechanisms differ from the positive-Kerr, Kac-scaled Hamiltonian confinement studied here. Other Kerr-resonator stabilization schemes are discussed in Refs.~\cite{Mirrahimi2014,Puri2017}.

\section{One-photon benchmark and the origin of the discontinuous onset}
\label{sec:SM_one_photon}

The ordinary one-photon Dicke model is the natural benchmark for identifying which features are specific to collective pair amplification. For completeness, consider
\begin{equation}
  H_{\rm 1ph} = \omega_c\hat n+\omega_0J_z + \frac{\lambda_1}{\sqrt N}(a+a^\dagger)J_x + \frac{\chi}{2N}\hat n^2 .    
\end{equation}
Using the photon density $n$ and the coupling conventions of the main text, spin minimization for a collective one-photon coupling $\lambda_1$ gives
\begin{equation}
 e_{\rm 1ph}(n) = \omega_c n+\frac{\chi}{2}n^2 -\frac12\sqrt{\omega_0^2+4\lambda_1^2n}.
 \label{eq:SM_one_photon_energy}
\end{equation}
Near the normal (\(n=0\)) boundary,
\begin{equation}
 e_{\rm 1ph}(n) = -\frac{\omega_0}{2} + \Big( \omega_c-\frac{\lambda_1^2}{\omega_0} \Big)n + O(n^2).
 \label{eq:SM_one_photon_small_n}
\end{equation}
The one-photon interaction therefore changes the coefficient linear in $n$ and destabilizes the normal state continuously at \(\lambda_{1,c}=\sqrt{\omega_c\omega_0}\). 
The Kerr term affects the nonlinear saturation but not this threshold.  In the two-photon model, by contrast, the coupling first changes the energy at order $n^2$, while $\left.\partial_ne\right|_{n=0}=\omega_c>0$.  

The large-field (\(n\rightarrow\infty\)) behavior is equally different.  For the bare one-photon model ($\chi=0$),
\begin{equation}
 e_{\rm 1ph}(n) = \omega_c n-\lambda_1\sqrt n+O(n^{-1/2}).
\end{equation}
Thus, the positive cavity term already confines the spectrum.  Pair amplification instead produces the asymptotic competition $e(n)\sim(\omega_c-\lambda)n$ when $\chi=0$ and therefore requires an independent large-field regularizer.

The two ordered phases consequently differ in both symmetry and onset.  
The one-photon Dicke transition has a continuous $\mathbb Z_2$-symmetry-breaking onset, whereas the Kerr-confined two-photon model has a discontinuous onset with four symmetry-related bright configurations.


\section{Conventions, scaling, and \texorpdfstring{finite-$N$}{finite-N} confinement}
\label{sec:SM_operator_boundedness}
\label{sec:SM_conventions}
Throughout the SM, \(H\) denotes the Hamiltonian in Eq.~\eqref{eq:Hmain}, with \(\omega_c,\omega_0>0\) and \(\chi,\lambda\geq0\), all held fixed as \(N\to\infty\). 
Here we show finite-\(N\) lower boundedness and test the bare-model limit and normal-ordering correction.

\subsection{\texorpdfstring{Finite-$N$}{Finite-N} operator lower bound}

Using the following identities of the cavity operators
\begin{equation}
 2\hat n+1\pm\left(a^2+a^{\dagger2}\right) =
 \begin{cases}
 (a+a^\dagger)^2,&+,\\
 [i(a^\dagger-a)]^2,&-,
 \end{cases}
 \geq0,
\end{equation}
we can write 
\begin{equation}
 -\left(2\hat n+1\right)  \leq  a^2+a^{\dagger2}  \leq  2\hat n+1.
 \label{eq:SM_a2_bound}
\end{equation}
This bound applies separately in each \(J_x\) eigenspace. Using \(\|J_x\|\leq N/2\), Eq.~\eqref{eq:SM_a2_bound} gives
\begin{equation}
 \frac{\lambda}{N} \left(a^2+a^{\dagger2}\right)J_x  \geq  -\frac{\lambda}{2}(2\hat n+1)=  -\lambda\hat n-\frac{\lambda}{2}.
 \label{eq:SM_interaction_lower_bound}
\end{equation}
Combining this with \(J_z\geq-N/2\) in Eq.~\eqref{eq:Hmain}, we obtain
\begin{equation}
  H\geq F(\hat n)-\frac{\omega_0N}{2}-\frac{\lambda}{2}\,,
  \label{sq:SM_H_inequality_1}
\end{equation}
where \(F(x)=\frac{\chi}{2N}x^2+(\omega_c-\lambda)x\),
with \(x\geq0\). 
Since photon-number eigenvalues are nonnegative integers, minimizing \(F(x)\) over all real \(x\geq0\) gives a valid lower bound for \(F(\hat n)\). For \(\lambda\leq\omega_c\), both terms are nonnegative, so \(F(\hat n)\geq0\). For \(\lambda>\omega_c\), completing the square gives
\(  F(\hat n) = \frac{\chi}{2N} \big[\hat n-\frac{N(\lambda-\omega_c)}{\chi}\big]^2 -\frac{N(\lambda-\omega_c)^2}{2\chi}\). Therefore,
\begin{align}
  F(\hat n) 
            &\geq-\frac{N(\lambda-\omega_c)^2}{2\chi}.
  \label{eq:SM_Fn_inequality}
\end{align}
Combining the two cases, Eqs.~\ref{sq:SM_H_inequality_1} and \eqref{eq:SM_Fn_inequality}, we have
\begin{equation}
H\geq
-\frac{N}{2\chi}(\lambda-\omega_c)_+^2
-\frac{\omega_0N}{2}-\frac{\lambda}{2},
\label{eq:SM_final_lower_bound}
\end{equation}
where \(x_+=\max(x,0)\). Thus \(H\) is bounded from below for every finite \(N\) and coupling \(\lambda\) with \(\chi>0\). This holds even without a CS or large-\(N\) approximation.

The above estimates also establish a self-adjoint Hamiltonian \(H\)  on the full spin--photon Hilbert space.  Write 
\begin{equation}
  H_K=\frac{\chi}{2N}\hat n^2, \qquad
  V=\frac{\lambda}{N}(a^2+a^{\dagger2})J_x.
\end{equation}
For any normalized state with finite \(\langle\hat n^2\rangle\), Eq.~\eqref{eq:SM_a2_bound} and \(\|J_x\|\leq N/2\) give 
\begin{equation}
    |\langle V\rangle| \leq\lambda\left(\langle\hat n\rangle+\frac12\right).
\end{equation}
Thus, the first moment of \(\hat n\) controls the interaction energy, while the second moment controls the Kerr energy.

For any \(\epsilon>0\), positivity of \(\epsilon(\hat n-\frac{1}{2\epsilon})^2\) implies 
\begin{equation}
    \hat n\leq\epsilon\hat n^2+\frac{1}{4\epsilon}.
    \label{eq:SM_epsilon_inequality}
\end{equation}
Consequently,
\begin{align}
  |\langle V\rangle| \leq 
  \frac{2N\lambda\epsilon}{\chi}\langle H_K\rangle + \frac{\lambda}{4\epsilon}+\frac{\lambda}{2}.
\end{align}
For \(\lambda>0\), choosing \(\epsilon=\eta\chi/(2N\lambda)\) gives, for any \(\eta>0\),
\begin{equation}
  |\langle V\rangle| \leq\eta\langle H_K\rangle+C_{\eta,N}\,,
  \label{eq:SM_interaction_form_bound}
\end{equation}
where \( C_{\eta,N}=\frac{N\lambda^2}{2\eta\chi}+\frac{\lambda}{2}\) is a state-independent constant. The case \(\lambda=0\) is immediate as \(V=0\).
Note that \(\eta\) can be arbitrarily small, at the cost of increasing \(C_{\eta,N}\). The Kerr term \(H_K\) has a closed positive quadratic form on the states with finite \(\langle\hat n^2\rangle\).  The cavity term \(\omega_c\hat n\) is controlled by the same inequality (Eq.~\eqref{eq:SM_epsilon_inequality}), while \(\omega_0J_z\) is bounded at fixed \(N\). Standard quadratic-form theory thus yields a closed, lower-bounded energy form and its associated self-adjoint Hamiltonian~\cite{Teschl2014QM}. Thus, the model has both a finite lower energy bound and well-defined unitary quantum dynamics.

\subsection{Bare-model limit and normal ordering}
\label{subsec:normal_order}

Applying the interaction bound in Eq.~\eqref{eq:SM_interaction_lower_bound} and the spin bound \(J_z\geq-N/2\) directly to $H_{\rm bare}$ in Eq.~\eqref{eq:bare} gives
\begin{equation}
  H_{\rm bare} \geq (\omega_c-\lambda)\hat n - \frac{\lambda}{2}-\frac{\omega_0N}{2}.
  \label{eq:SM_bare_lower_bound}
\end{equation}
Let \(\mathcal E_0(N)\)
denote the lowest spectral energy of \( H_{\rm bare}\).
The cavity vacuum with all \(N\) emitters polarized along \(-z\) has energy expectation \(-\frac{\omega_0N}{2}\). Thus \(\frac{\mathcal E_0(N)}{N} \leq - \frac{\omega_0}{2}\). For \(\lambda\leq\omega_c\), since \((\omega_c-\lambda)\hat n\) is nonnegative, Eq.\eqref{eq:SM_bare_lower_bound} gives \( -\frac{\omega_0}{2}-\frac{\lambda}{2N} \leq \frac{\mathcal E_0(N)}{N} \). Together,
\begin{equation}
  -\frac{\omega_0}{2}-\frac{\lambda}{2N} \leq \frac{\mathcal E_0(N)}{N} \leq - \frac{\omega_0}{2}.
  \label{eq:SM_bare_energy_density_sandwich}
\end{equation}
For fixed \(\lambda<\omega_c\), combining the same bounds (Eq.~\eqref{eq:SM_bare_lower_bound} and Eq.~\eqref{eq:SM_bare_energy_density_sandwich}) also gives, in every ground state,
\begin{equation}
  \frac{\langle\hat n\rangle}{N} \leq \frac{\lambda}{2N(\omega_c-\lambda)} \longrightarrow0.
  \label{eq:SM_bare_density_vanishes}
\end{equation}
Thus, the confined bare regime has energy density \(-\omega_0/2\) and vanishing photon density. The energy-density limit also holds at \(\lambda=\omega_c\), but the density estimate does not.

To expose the loss of confinement, introduce \(Q=(a+a^\dagger)/\sqrt2\) and \(P=(a-a^\dagger)/(i\sqrt2)\). Taking the spin expectation in the trial state with \(J_x=-N/2\), for which \(\langle J_z\rangle=0\), leaves the field operator
\begin{equation}
h_{\rm bare} = \frac{\omega_c-\lambda}{2}Q^2 + \frac{\omega_c+\lambda}{2}P^2 -\frac{\omega_c}{2}.
\label{eq:SM_quadrature_bare}
\end{equation}
This is a variational channel, not an invariant spin block of \(H_{\rm bare}\). 
At \(\lambda=\omega_c\), the \(Q^2\) stiffness vanishes. Squeezed trial states with \(\langle Q^2\rangle\to\infty\) and \(\langle P^2\rangle\to0\) have finite limiting energy, exhibiting the flat channel associated with spectral collapse~\cite{Felicetti2015,Lo2021Multiqubit}. For \(\lambda>\omega_c\), the same sequence gives \(\langle h_{\rm bare}\rangle\to-\infty\), proving unboundedness from below.

Positive Kerr reverses this large-occupation instability. Using \(2\hat n=Q^2+P^2-1\) and
\(\langle\hat n^2\rangle\geq\langle\hat n\rangle^2\),
\begin{equation}
 \langle H_K\rangle \geq\frac{\chi}{8N} \bigl(\langle Q^2\rangle-1\bigr)^2,
\end{equation}
with \(\langle Q^2\rangle\geq1\). Its growth therefore dominates the negative contribution linear in \(\langle Q^2\rangle\).

A physical Kerr interaction is often normally ordered as \(\chi a^{\dagger2}a^2/(2N)\).  Since \(a^{\dagger2}a^2=\hat n^2-\hat n\), the corresponding Hamiltonian satisfies
\begin{equation}
  H^{\rm n.o.} \geq \frac{\chi}{2N}\hat n^2 +\left(\omega_c-\lambda-\frac{\chi}{2N}\right)\hat n -\frac{\lambda}{2}-\frac{\omega_0N}{2}.
  \label{eq:SM_normal_ordered_lower_bound}
\end{equation}
Normal ordering therefore preserves lower boundedness and the thermodynamic energy functional but introduces the finite-\(N\) cavity-frequency shift \(-\chi/(2N)\). This shift affects finite-size branch-crossing couplings unless compensated by the cavity detuning, as discussed in Sec.~\ref{sec:SM_circuit_QED}.

\section{Exact thermodynamic limit and marginal Kac scaling}
\label{sec:SM_exact_thermodynamic_limit}

Here, we establish the exact thermodynamic ground-state energy density oof \(H\) in Eq.~\eqref{eq:Hmain} for $\chi>0$. First, we reduce the ground-state problem to the maximal-spin sector. Covariant and contravariant CS symbols then give matching upper and lower bounds, and Kerr coercivity controls their convergence. Finally, we identify the marginal scaling of the physical Kerr coefficient.

The collective-spin operators \(J_\mu=\frac12\sum_{k=1}^N\sigma_\mu^{(k)}\) define \(J^2=\sum_{\mu=x,y,z}J_\mu^2\), with \(J^2|j,m\rangle=j(j+1)|j,m\rangle\) and \(J_z|j,m\rangle=m|j,m\rangle\), where \(m=-j,\ldots,j\). Since \([H,J^2]=0\), different total-spin sectors decouple and can be diagonalized independently. For \(N\) spin-\(\tfrac12\) emitters, there are \(\lfloor N/2\rfloor+1\) distinct total-spin values, with \(j=N/2,N/2-1,\ldots,j_{\min}\), where \(j_{\min}=0\) (\(\frac12\)) for even (odd) \(N\). Among these, the maximal-spin sector \(j=N/2\) is the symmetric Dicke sector, spanned by \(N+1\) states with \(r=0,\ldots,N\) excitations. It has the form
\begin{equation}
  |D_N^{(r)}\rangle = \frac{1}{\sqrt{\binom{N}{r}}} \sum_{\substack{s_1,\ldots,s_N\in\{g,e\}\\ 
  \#\{i:s_i=e\}=r}} |s_1,\ldots,s_N\rangle.
\end{equation}
The sum includes all distinct arrangements of \(r\) excited and \(N-r\) ground-state emitters, while the full emitter Hilbert space has \(2^N\) states. 
Several independent copies of a given spin-\(j\) multiplet may occur.
Since \(H\) acts identically on each copy, we suppress the multiplicity index.

\subsection{Exact reduction to the maximal-spin sector}
\label{sec:SM_maximal_spin}

Let \(E_{N,j}\) be the lowest energy in a spin-\(j\) sector for fixed \(N\), so that \(E_0(N)=\min_j E_{N,j}\). Although \([H,J^2]=0\) ensures sector decoupling, it does not identify which sector contains a ground state. To determine this, we compare the minimum energies of successive allowed spin sectors \(j\) and \(j+1\), keeping \(N\) fixed. We show below that \(E_{N,j+1}\leq E_{N,j}\) for \(j<N/2\), establishing the exact ground-state reduction to \(j=N/2\).

In the photon-number basis \(|n_a\rangle\), introduce the unitary \(U_b=\exp\!\big[\frac{i\pi}{2}\hat n(\hat n-1)\big]\), such that \(U_b|n_a\rangle=(-1)^{\lfloor n_a/2\rfloor}|n_a\rangle\). Since \(U_b(a^2+a^{\dagger2})U_b^\dagger =-(a^2+a^{\dagger2})\), the transformed Hamiltonian \(\widetilde H=U_bHU_b^\dagger\) preserves the spectrum and total-spin sector decomposition, while making all off-diagonal matrix elements nonpositive  \((\lambda\geq0)\). Explicitly, in the joint basis \(|n_a;j,m\rangle=|n_a\rangle\otimes|j,m\rangle\), the nonzero off-diagonal matrix elements are
\begin{align}
  \langle n_a+2;j,m\pm1|\widetilde H|n_a;j,m\rangle = -\frac{\lambda}{N} p_{n_a}c_{j,m}^{\pm}\leq0,
  \label{eq:SM_negative_offdiagonal}
\end{align}
where \(c_{j,m}^{\pm}=\frac12\sqrt{j(j+1)-m(m\pm1)}\) are the spin-flip amplitudes of \(J_x\), and \(p_{n_a}=\sqrt{(n_a+1)(n_a+2)}\).
The reverse transitions follow by Hermitian conjugation. All other off-diagonal elements vanish.

At finite photon cutoff \(M\), a normalized ground state of the restricted Hamiltonian in sector \(j\) has the expansion
\begin{equation}
  |\widetilde\psi_j^{(M)}\rangle = \sum_{n_a=0}^{M}\sum_{m=-j}^{j} A_{n_a,m}^{(j)}|n_a;j,m\rangle.
\end{equation}
Equivalently, \(|\widetilde\psi_j^{(M)}\rangle=\sum_\xi A_\xi|\xi;j\rangle\), where \(\xi=(n_a,m)\), \(|\xi;j\rangle=|n_a;j,m\rangle\), and \(A_\xi=A_{n_a,m}^{(j)}\). Its energy is \(E_{N,j}^{(M)}\). For \(\xi\neq\xi'\), the real matrix elements \(\widetilde H_{\xi\xi'}=\langle\xi;j|\widetilde H|\xi';j\rangle\leq0\). Since \(\operatorname{Re}(A_\xi^*A_{\xi'})\leq|A_\xi||A_{\xi'}|\),
\begin{equation}
\widetilde H_{\xi\xi'}|A_\xi||A_{\xi'}|
\leq\widetilde H_{\xi\xi'}
\operatorname{Re}(A_\xi^*A_{\xi'}).
\end{equation}
Thus, replacing each coefficient by its absolute value preserves normalization and diagonal energy, while the off-diagonal energy cannot increase. The resulting state \(|\widetilde\psi_{\rm abs}\rangle = \sum_\xi|A_\xi||\xi;j\rangle\)  satisfies \(\langle\widetilde\psi_{\rm abs}|\widetilde H| \widetilde\psi_{\rm abs}\rangle\leq E_{N,j}^{(M)}\). The variational principle gives the reverse inequality, so equality holds and we may choose \(A_{n_a,m}^{(j)}\geq0\).

To compare successive allowed spin sectors at fixed \(N\), let $j<N/2$ and construct the normalized trial state
\begin{equation}
  |\widetilde\phi_{j+1}^{(M)}\rangle = \sum_{n_a=0}^{M}\sum_{m=-j}^{j} A_{n_a,m}^{(j)}|n_a;j+1,m\rangle,
\end{equation}
with \(\langle n_a;j+1,\pm(j+1)
\mid\widetilde\phi_{j+1}^{(M)}\rangle=0\) and \(n_a=0,\ldots,M\),
preserving normalization and diagonal energy. For spin-flip transitions shared by both sectors, \(c_{j+1,m}^{\pm}\geq c_{j,m}^{\pm}\), so the corresponding off-diagonal elements in Eq.~\eqref{eq:SM_negative_offdiagonal} become more negative or remain unchanged. Since the coefficients are nonnegative, the variational principle gives
\begin{equation}
  E_{N,j+1}^{(M)} \leq \langle\widetilde\phi_{j+1}^{(M)}| \widetilde H|\widetilde\phi_{j+1}^{(M)}\rangle \leq E_{N,j}^{(M)}.
\end{equation}
The finite-photon cutoff spaces are nested, \( \mathcal H_j^{(M)} \subset\mathcal H_j^{(M+1)} \subset\mathcal H_j\). Consequently, increasing \(M\) introduces additional trial states, so the variational principle gives \(E_{N,j}^{(M+1)}\leq E_{N,j}^{(M)}\).At fixed $N,j$, the pair interaction is infinitesimally form-bounded by the Kerr term, which controls $\langle\hat n^2\rangle$. Thus, the union of finite-photon cutoff spaces is a quadratic-form core for the semibounded Hamiltonian [Eq.~\eqref{eq:SM_interaction_form_bound}], giving \(\lim_{M\to\infty}E_{N,j}^{(M)}=E_{N,j}\). Taking this limit in the finite-cutoff inequality gives
\begin{equation}
  E_{N,j+1}\leq E_{N,j}.
\label{eq:SM_exact_maximal_spin}
\end{equation}
Iterating over the allowed spin sectors yields
$E_0(N)=E_{N,N/2}$, so the symmetric Dicke sector gives the exact ground-state energy \( E_0(N)\) for every finite $N$.

\subsection{Coherent-state (CS) upper and lower symbols}
\label{sec:SM_exact_symbols}

Having established the exact reduction of the ground-state problem to $j=N/2$, we now bound its energy using bosonic and spin CSs. Their expectation values (covariant symbols) provide a variational upper bound, while their operator representations (contravariant symbols) yield a corresponding lower bound~\cite{Lieb1973Classical,LiebSeiringerYngvason2005}.

Let \(|z\rangle\) be a bosonic CS, satisfying \(a|z\rangle=z|z\rangle\) and \(\mathbf 1_b=\int\frac{d^2z}{\pi}|z\rangle\langle z|\). The bosonic covariant symbols \(O_{\rm cov}(z)=\langle z|\hat O|z\rangle\) are \(\hat n \leftrightarrow |z|^2,\,\, \hat n^2\leftrightarrow |z|^4+|z|^2,\,\, a^2+a^{\dagger2}\leftrightarrow z^2+\bar z^2\). The corresponding contravariant symbols, defined by \(\hat O=\int\frac{d^2z}{\pi}\,O_{\rm con}(z) |z\rangle\langle z|\), are
\[
\hat n \leftrightarrow |z|^2-1,\, 
\hat n^2 \leftrightarrow |z|^4-3|z|^2+1,\,\,  a^2+a^{\dagger2}
\leftrightarrow z^2+\bar z^2.
\]
Rescaling \(z=\sqrt N\,\alpha\), with \(\alpha=\sqrt n\,e^{i\phi}\), defines the intensive variables \(n=|\alpha|^2\) and \(\zeta=\alpha^2+\bar\alpha^2=2n\cos2\phi\).

Let \(|\boldsymbol\Omega;j\rangle=R|j,j\rangle\) be a spin CS oriented along the unit vector \(\boldsymbol\Omega=(\sin\theta\cos\varphi_s, \sin\theta\sin\varphi_s,\cos\theta)\). Here \(\theta\) and \(\varphi_s\) are the spin polar and azimuthal angles, and \(R=e^{-i\phi_sJ_z}e^{-i\theta J_y}\). 
It satisfies \((\boldsymbol\Omega\cdot\mathbf J) |\boldsymbol\Omega;j\rangle = j|\boldsymbol\Omega;j\rangle\), 
\( \mathbf J^2|\boldsymbol\Omega;j\rangle = j(j+1)|\boldsymbol\Omega;j\rangle \), and
\(\mathbf 1_j=(2j+1)\int\frac{d\boldsymbol\Omega}{4\pi} |\boldsymbol\Omega;j\rangle \langle\boldsymbol\Omega;j|\),
where \(d\boldsymbol\Omega = \sin\theta\,d\theta\, d\varphi_s\) is the solid-angle measure.
For \(\mu=x,y,z\), the covariant and contravariant spin symbols are
\((J_\mu)_{\rm cov}=j\Omega_\mu\) and \((J_\mu)_{\rm con}=(j+1)\Omega_\mu\), respectively, where \((J_\mu)_{\rm cov} =\langle\boldsymbol\Omega;j|J_\mu| \boldsymbol\Omega;j\rangle\), while the contravariant symbol enters 
through
\begin{equation}
  J_\mu=(2j+1)\int\frac{d\boldsymbol\Omega}{4\pi}\, (J_\mu)_{\rm con} |\boldsymbol\Omega;j\rangle \langle\boldsymbol\Omega;j|.
  \label{eq:SM_coherent_resolutions}
\end{equation}

Using the product CSs $|\sqrt{N}\alpha,\boldsymbol\Omega\rangle = |\sqrt{N}\alpha\rangle\otimes|\boldsymbol\Omega;N/2\rangle$, both the covariant and contravariant energy-density symbols share the leading term
\begin{equation}
  h(\alpha,\boldsymbol\Omega) = \omega_c n + \frac{\chi}{2}n^2 + \frac12\left(\omega_0\Omega_z+\lambda\zeta\Omega_x\right).
  \label{eq:SM_classical_symbol}
\end{equation}
This describes the classical field--spin energy before minimizing over the spin orientation.

The exact covariant energy symbol is \(H_{\rm cov}(\alpha,\boldsymbol\Omega) =\langle\sqrt N\alpha,\boldsymbol\Omega| H |\sqrt N\alpha,\boldsymbol\Omega\rangle\), giving the energy density
\begin{equation}
  \frac{H_{\rm cov}(\alpha,\boldsymbol\Omega)}{N} = h(\alpha,\boldsymbol\Omega) +\frac{\chi n}{2N}.
  \label{eq:SM_covariant_symbol_exact}
\end{equation}
The finite-\(N\) correction follows from \(\langle\hat n^2\rangle=N^2n^2+Nn\). By the variational principle, \(E_0(N)\leq H_{\rm cov}(\alpha,\boldsymbol\Omega)\) for every product CS, yielding the upper bound
\begin{equation}
  \frac{E_0(N)}{N} \leq\inf_{\alpha,\boldsymbol\Omega} \left[h(\alpha,\boldsymbol\Omega) + \frac{\chi n}{2N}\right].
  \label{eq:SM_covariant_upper_bound}
\end{equation}

Unlike the covariant symbol, in the maximal-spin sector, the contravariant energy symbol \(H_{\rm con}(\alpha,\boldsymbol\Omega)\) reconstructs \(H\) through the exact operator representation (as understood in the quadratic-form sense)
\begin{equation}
  H=\int d\nu_N^{(0)}\, H_{\rm con}(\alpha,\boldsymbol\Omega)\, |\sqrt N\alpha,\boldsymbol\Omega\rangle \langle\sqrt N\alpha,\boldsymbol\Omega|,
  \label{eq:SM_P_representation}
\end{equation}
where \(d\nu_N^{(0)}= N(2j+1)\, \frac{d^2\alpha\, d\boldsymbol\Omega}{4\pi^2}\). Using the contravariant symbols defined above, we obtain (for \(j=N/2\))
\begin{align}
\frac{H_{\rm con}(\alpha,\boldsymbol\Omega)}{N}
={}&h(\alpha,\boldsymbol\Omega)
-\frac{\omega_c}{N}
-\frac{3\chi n}{2N}
+\frac{\chi}{2N^2}
\nonumber\\
&+\frac{\omega_0\Omega_z+\lambda\zeta\Omega_x}{N}.
\label{eq:SM_contravariant_symbol_exact}
\end{align}

For any normalized finite-energy state \(|\Psi\rangle\) in the maximal-spin sector, define the nonnegative overlap weight \(w_\Psi(\alpha,\boldsymbol\Omega) = |\langle\sqrt N\alpha,\boldsymbol\Omega |\Psi\rangle|^2\) with \(\int w_\Psi\,d\nu_N^{(0)}=1\).
Thus, these weights behave like a normalized probability distribution.
Equation~\eqref{eq:SM_P_representation} therefore gives
\begin{align} 
  \langle\Psi|H|\Psi\rangle &=\int d\nu_N^{(0)}\,w_\Psi\,H_{\rm con}(\alpha,\boldsymbol\Omega) \nonumber\\
  &\geq \inf_{\alpha,\boldsymbol\Omega}H_{\rm con}(\alpha,\boldsymbol\Omega).
  \label{eq:SM_P_symbol_lower_bound}
\end{align}
Since this holds for arbitrary states, including
squeezed or spin--field-entangled states, the exact maximal-spin reduction
gives
\begin{align}
    \frac{E_0(N)}{N} \geq \inf_{\alpha,\boldsymbol\Omega} &\left[h(\alpha,\boldsymbol\Omega)
-\frac{\omega_c}{N}
-\frac{3\chi n}{2N}
+\frac{\chi}{2N^2}\right.\nonumber\\
&\qquad\qquad\left.+\frac{\omega_0\Omega_z+\lambda\zeta\Omega_x}{N}
\right].
  \label{eq:SM_covariant_lower_bound}
\end{align}

\subsection{Spin and phase minimization}
\label{sec:SM_spin_phase_minimization}
To obtain the spin-minimized functional $e(n,\phi)=\min_{\boldsymbol\Omega^2=1} h(\alpha,\boldsymbol\Omega)$, write the spin-dependent term as
\begin{equation}
  \frac12(\omega_0\Omega_z+\lambda\zeta\Omega_x) = \frac12\boldsymbol B_{\rm eff}\cdot\boldsymbol\Omega,
  \label{eq:SM_effective_spin_field}
\end{equation}
where \(\boldsymbol B_{\rm eff} = (2\lambda n\cos2\phi,0,\omega_0)\). For fixed \(n\) and \(\phi\), the energy is minimized when the collective spin is antiparallel to \(\boldsymbol B_{\rm eff}\), i.e., \(\boldsymbol\Omega = -\boldsymbol B_{\rm eff}/|\boldsymbol B_{\rm eff}|\). Thus,
\begin{align}
  \min_{\boldsymbol\Omega^2=1} \frac12\boldsymbol B_{\rm eff}\cdot\boldsymbol\Omega
  &=-\frac12\sqrt{ \omega_0^2+4\lambda^2n^2\cos^2(2\phi)}.
  \label{eq:SM_spin_minimization}
\end{align}

Substituting this result into Eq.~\eqref{eq:SM_classical_symbol} gives the spin-minimized energy density
\begin{equation}
 e(n,\phi) = \omega_cn+\frac{\chi}{2}n^2 - \frac12\sqrt{\omega_0^2+4\lambda^2n^2\cos^2(2\phi)}.
 \label{eq:SM_phase_energy}
\end{equation}
For \(\lambda\, n>0\), minimizing over the field phase maximizes \(\cos^2(2\phi)\), selecting \(|\cos2\phi|=1\) and the four symmetry-related phases \(\phi=0,\pi/2,\pi,3\pi/2\). Defining \(R(n)=\sqrt{\omega_0^2+4\lambda^2n^2}\), the resulting energy density is
\begin{equation}
  e(n)=\omega_cn+\frac{\chi}{2}n^2-\frac12R(n).
  \label{eq:SM_en}
\end{equation}
At fixed $n$, the covariant and contravariant symbols have the same minimizing spin orientation and field phases. 
Evaluating these minimizations in Eqs.~\eqref{eq:SM_covariant_upper_bound} and \eqref{eq:SM_covariant_lower_bound} gives the finite-\(N\) energy density bounds
\begin{align}
  e_N^{+}(n) &=e(n)+\frac{\chi n}{2N},
  \label{eq:SM_exact_symbol_bound_plus}\\
  e_N^{-}(n) &=e(n)-\frac{R(n)+\omega_c+3\chi n/2}{N} +\frac{\chi}{2N^2}.
  \label{eq:SM_exact_symbol_bound_minus}
\end{align}

\subsection{Convergence and concentration of exact ground states}
\label{sec:SM_exact_convergence}

We now show that both covariant and contravariant bounds converge to the same ground-state energy density, \(e_\infty\!=\min_{n\geq0}e(n)\). Using \(R(n)\!\leq\!\omega_0+2\lambda n\), we obtain
\begin{equation}
  e(n)\geq \frac{\chi}{2}n^2 + (\omega_c-\lambda)n-\frac{\omega_0}{2}.
  \label{eq:SM_e_coercive_bound}
\end{equation}
 For fixed $\chi>0$, the positive quadratic term $\chi n^2/2$ eventually dominates the negative contributions, which grow at most linearly with $N$-independent coefficients. Each minimum cannot exceed the corresponding value at $n=0$. Since these reference values remain bounded as $N$ varies, all minimizers lie in a common $N$-independent interval. Throughout this interval, $e_N^\pm(n)-e(n)=O(N^{-1})$ uniformly. The energy bounds then give
\begin{equation}
  \left|\frac{E_0(N)}{N}-e_\infty\right|\leq\frac{C}{N},
  \label{eq:SM_exact_TDL_theorem}
\end{equation}
where $C$ is independent of $N$. This establishes Eq.~\eqref{eq:main_exact_thermodynamic_limit}

The same bounds also determine where the exact ground states concentrate in the thermodynamic limit. For a normalized ground state \(|\Psi_N\rangle\) in the maximal-spin sector, define the scaled Husimi probability measure
\begin{equation*}
  d\nu_N(\alpha,\boldsymbol\Omega) = |\langle\sqrt N\alpha,\boldsymbol\Omega |\Psi_N\rangle|^2\,d\nu_N^{(0)} \equiv w_\Psi(\alpha,\boldsymbol\Omega)\,d\nu_N^{(0)},
\end{equation*}
with \(\int d\nu_N=1\).
We first control the weight at large field amplitudes. Combining Eq.~\eqref{eq:SM_interaction_lower_bound} with the vacuum trial bound \(E_0(N)\leq-\omega_0N/2\) gives
\begin{equation}
  \frac{\langle\hat n^2\rangle}{N^2} \leq \frac{4(\lambda-\omega_c)_+^2}{\chi^2} + \frac{2\lambda}{\chi N}.
\end{equation}
Here we used
\((\lambda-\omega_c)_+\langle\hat n\rangle \leq \frac{\chi}{4N}\langle\hat n^2\rangle + \frac{N(\lambda-\omega_c)_+^2}{\chi}\).
Consequently,
\begin{equation}
  \int|\alpha|^4\,d\nu_N = \frac{\langle\hat n^2+3\hat n+2\rangle}{N^2} = O(1).
\label{eq:SM_Husimi_fourth_moment}
\end{equation}
with expectations taken in \(|\Psi_N\rangle\). This moment bound prevents probability from escaping to arbitrarily large field amplitudes (tightness). 

The symbol corrections grow at most linearly in $|\alpha|^2$, so the same moment bound makes their average $O(N^{-1})$. The exact contravariant representation therefore gives
\begin{align*}
  \frac{E_0(N)}{N} =\!\int\frac{H_{\rm con}(\alpha,\boldsymbol\Omega)}{N}\,d\nu_N
  = \!\int\! h(\alpha,\boldsymbol\Omega)\,d\nu_N+O(N^{-1}).
\end{align*}
Combining this identity with Eq.~\eqref{eq:SM_exact_TDL_theorem} yields
\begin{align}
  \int \left[h(\alpha,\boldsymbol\Omega) -e_\infty\right]\,d\nu_N=O(N^{-1}).
\end{align}
Since $h-e_\infty\geq0$, the Husimi probability weight of field amplitudes $\alpha$ and spin orientations $\boldsymbol\Omega$ whose energy density exceeds the global minimum $e_\infty$ by at least a fixed $\epsilon>0$ obeys
\begin{equation*}
  \nu_N\bigl(\{h-e_\infty\geq\epsilon\}\bigr)
  \leq\frac{1}{\epsilon}\int(h-e_\infty)\,d\nu_N
  =O(N^{-1}).
\end{equation*}
Thus, every weak limit assigns probability one to the global minimizers of $h$. At coexistence, these bounds leave the relative weights of the degenerate minima undetermined.


\subsection{Proof of the marginal Kerr scaling}
\label{sec:SM_Kac_trichotomy}
Positive Kerr confines every finite system, but its thermodynamic effect depends on how its coefficient scales with $N$.
At fixed $\omega_c,\omega_0>0$ and $\lambda\geq0$, let $H_N(K_N)=H_{\rm bare}+H_K$, where \(H_K=\frac{K_N}{2}\hat n^2 =\frac{\chi_N}{2N}\hat n^2\),  with \(\chi_N=\chi=NK_N>0\). We denote its ground-state energy by $E_0(N;K_N)$, abbreviated as $E_0(N)$.
At fixed $n$, the leading Kerr energy density is $\chi_N n^2/2$. Equation~\eqref{eq:SM_interaction_lower_bound} gives
\begin{equation}
  H_N(K_N)\geq \frac{K_N}{2}\hat n^2
  +(\omega_c-\lambda)\hat n
  -\frac{\lambda}{2}-\frac{\omega_0N}{2}.
  \label{eq:SM_general_Kerr_lower_bound}
\end{equation}
Together with the vacuum bound $E_0(N)\leq-\omega_0N/2$, this distinguishes three regimes.
Write $d=(\lambda-\omega_c)_+$.

\emph{(i) $\chi_N\to\infty$: vanishing photon density.}
Minimizing the lower-bound polynomial gives
\begin{equation}
  -\frac{\omega_0}{2}
  -\frac{d^2}{2\chi_N}
  -\frac{\lambda}{2N}
  \leq \frac{E_0(N)}{N}
  \leq -\frac{\omega_0}{2}.
\end{equation}
Hence $E_0(N)/N\to-\omega_0/2$. Using $\langle\hat n^2\rangle\geq\langle\hat n\rangle^2$ in the same bound, the ground-state photon density satisfies
\begin{equation}
  \frac{\langle\hat n\rangle}{N}
  \leq \frac{d+\sqrt{d^2+K_N\lambda}}{NK_N}
  \longrightarrow0.
\end{equation}
Thus Kerr suppresses an extensive bright population. A fixed Kerr coefficient $K_N=K>0$ belongs to this regime: it confines the Hamiltonian at every finite $N$, but the thermodynamic photon density vanishes.

\emph{(ii) $\chi_N\to\chi\in(0,\infty)$:
finite Kerr competition.}
Both covariant and contravariant symbols converge to the same confined functional. Using arguments of Sec.~\ref{sec:SM_exact_convergence}, 
\begin{equation}
 \!\!\lim_{N\to\infty}\!\!\frac{E_0(N)}{N} \!=\! \min_{n\geq0}\! \left[ \omega_c n \!+\! \frac{\chi}{2}n^2 \!-\!\frac12\sqrt{\omega_0^2+4\lambda^2n^2} \right].
\end{equation}
Kerr confinement and pair amplification both remain finite in the energy per emitter, permitting a stable bright phase at nonzero photon density. This is the thermodynamic regime studied in the main text; whether the normal or bright phase is selected depends on $\lambda$.

\emph{(iii) $\chi_N\to0$: loss of a finite postcollapse energy density.}
For $\lambda>\omega_c$, choose a real bosonic CS with density $n_{\rm tr}=d/\chi_N$ and a spin state polarized along $-x$, with $J_x=-N/2$. Its energy yields the variational bound
\begin{equation}
  \frac{E_0(N)}{N}
  \leq -\frac{d^2}{2\chi_N}+\frac{d}{2N}
  \longrightarrow-\infty.
\end{equation}
Although every finite-$N$ Hamiltonian remains lower bounded, the Kerr coefficient decreases too rapidly to retain a finite ground-state energy density above the bare collapse threshold.

For a finite, nonzero limiting Kerr contribution at fixed nonzero photon density, the marginal scaling is
\begin{equation}
  K_N=\frac{\chi+o(1)}{N},
  \qquad 0<\chi<\infty.
  \label{eq:SM_unique_marginal_Kerr_scaling}
\end{equation}
This scaling permits, but does not require, a nonzero ground-state photon density.

\begin{figure*}[t]
\includegraphics[width=\textwidth,height=8cm]{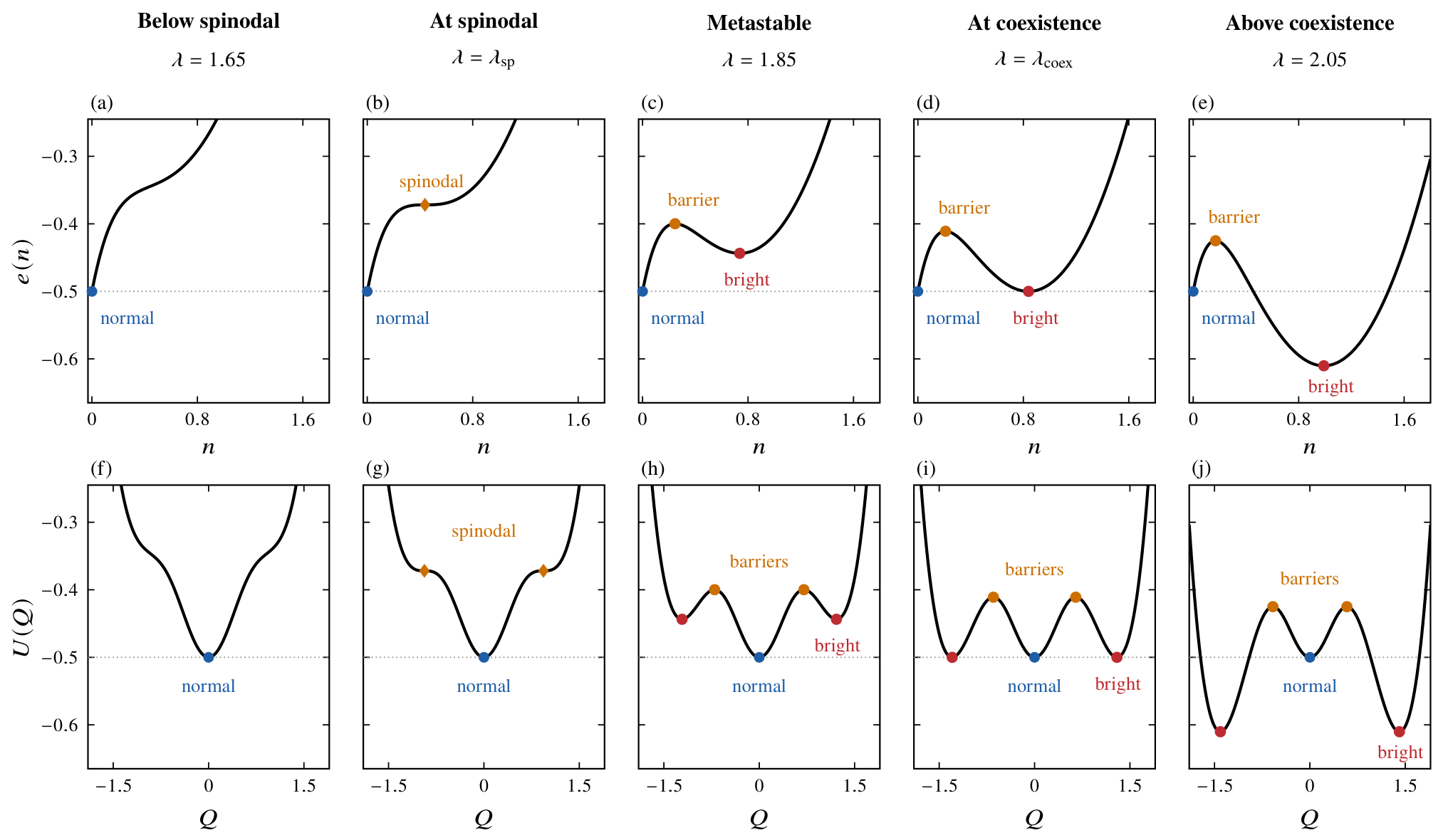}
\vspace{-3ex}
\caption{\textbf{Energy landscapes across the normal--bright transition.} (a)--(e) Photon-density landscape \(e(n)\). (f)--(j) Real-amplitude cuts \(U(Q)=e(Q^2/2)\), with the intensive coherent-state coordinate \(Q=\sqrt2\,\mathrm{Re}\alpha\). All panels use \(\omega_c=\omega_0=\chi=1\). Columns correspond to \(\lambda=1.65,\lambda_{\rm sp},1.85,\lambda_{\rm coex},2.05\), where \(\lambda_{\rm sp}\simeq1.724513\) and \(\lambda_{\rm coex}\simeq1.925325\). Blue and red circles mark normal and bright minima, orange circles mark barriers, and diamonds mark the marginal spinodal points. At the spinodal, the quadrature cut has stationary shoulders (inflection), not separate barriers and bright minima. At coexistence, the normal and bright minima are degenerate; above it, the normal minimum is metastable. Dotted lines show \(e(0)=-1/2\), with a common energy scale. When present, only the real-amplitude bright pair is visible. The other pair lies on the imaginary-amplitude axis.}
\label{fig:SM_energy_landscapes}
\end{figure*}

\section{Exact branch structure and global phase diagram}
\label{sec:SM_branch_selection}

For \(\omega_c,\omega_0,\chi>0\) and \(\lambda\geq0\), we now determine the stationary branches of \(e(n)\), their global selection, and the limiting ground-state observables.

\subsection{Stationary curve and saddle-node spinodal}
\label{subsec:SM_stationary_curve}

Recall Eq.~\eqref{eq:SM_en}, $e(n)=\omega_cn+\chi n^2/2-R(n)/2$, where $R(n)=\sqrt{\omega_0^2+4\lambda^2n^2}$. Writing $A(n)=\omega_c+\chi n$ gives $e'(n)=A(n)-2\lambda^2n/R(n)$. Since $e'(0)=\omega_c>0$, the normal boundary remains a local minimum at all \(\lambda>0\).

For $n>0$, stationarity requires $A(n)R(n)=2\lambda^2n$. Setting $y=\lambda^2$, we get $y^2-A^2y-A^2\omega_0^2/(4n^2)=0$. Selecting the positive root yields
\begin{equation}
  \lambda_{\rm st}^2(n) =\frac12\left[
    A^2(n)+A(n)\sqrt{A^2(n)+\frac{\omega_0^2}{n^2}} \,\right].
  \label{eq:SM_stationary_curve_compact}
\end{equation}
Note that $\lambda_{\rm st}^2(n)\sim\omega_c\omega_0/(2n)$ as $n\to0^+$ and $\lambda_{\rm st}^2(n)\sim\chi^2n^2$ as $n\to\infty$. These ensure an interior minimum.

Implicit differentiation of $e'(n;\lambda_{\rm st}(n))=0$ shows that extrema of this curve satisfy $e'(n)=e''(n)=0$. Using $e''(n)=\chi-2\lambda^2\omega_0^2/R^3(n)$, these conditions reduce to
\begin{equation}
  4\chi n^3(\omega_c+\chi n)^3=\omega_c^2\omega_0^2.
  \label{eq:SM_spinodal_unique_condition}
\end{equation}
The left-hand side increases strictly from zero to infinity, so there is exactly one positive solution,
\begin{equation}
  n_{\rm sp} =\frac{\sqrt{\omega_c^2+(4\chi\omega_c\omega_0)^{2/3}} -\omega_c}{2\chi}.
  \label{eq:SM_spinodal}
\end{equation}
Thus, the stationary curve has a unique global minimum at $n_{\rm sp}$, with $\lambda_{\rm sp}=\lambda_{\rm st}(n_{\rm sp})$ and $\lambda_{\rm sp}^2=(\omega_c+\chi n_{\rm sp})^3/\omega_c$.

No positive stationary root exists below $\lambda_{\rm sp}$. At $\lambda_{\rm sp}$, there is one marginal stationary point, where $e'(n_{\rm sp})\!=\!e''(n_{\rm sp})\!=\!0$ and \( e'''(n_{\rm sp})\!=\! 24\lambda_{\rm sp}^4\omega_0^2n_{\rm sp}/
  R^5(n_{\rm sp})>0\).
This inflection appears as two symmetry-related shoulders in the real-quadrature cut [Fig.~\ref{fig:SM_energy_landscapes}(b),(g)]. Above $\lambda_{\rm sp}$, there are two roots $n_-<n_{\rm sp}<n_+$. Since $e'(n)>0$ near zero and at large $n$, $n_-$ is a local maximum of $e(n)$ (the barrier), while $n_+$ is the bright local minimum.

Equivalently, \(n_\pm(\lambda)\) are the two positive roots of
\begin{equation}
  (\omega_c+\chi n)^2(\omega_0^2+4\lambda^2n^2)
  =4\lambda^4n^2.
  \label{eq:SM_stationary_density_quartic}
\end{equation}
Squaring the stationary equation adds no additional positive roots, since both sides are positive  \((n>0)\). Define \(B_\lambda =\lambda^2-\omega_c^2 -\frac{\chi^2\omega_0^2}{4\lambda^2}\),
and \(\theta_\lambda = \frac13\arccos\!\left[ \frac{27\chi^2\omega_0^2\omega_c^2} {2B_\lambda^3}-1 \right]\), and introduce \(u_j(\lambda)= \sqrt{\omega_c^2+\frac{2}{3}B_\lambda  \left[ 1+\cos\!\left(\theta_\lambda-\frac{2\pi j}{3}\right)\right]}\), with \(j=0,1,2\).
Using these, we obtain
\begin{equation}
n_\pm(\lambda)=\frac{u_0-\omega_c\pm(u_1-u_2)}{2\chi}.  
\end{equation}

\subsection{Coexistence and global branch selection}
\label{subsec:SM_coexistence_selection}

At \(\lambda=\lambda_{\rm sp}\),
\(e'(n;\lambda_{\rm sp})>0\) for
\(0\leq n<n_{\rm sp}\), implying
\(e(n_{\rm sp};\lambda_{\rm sp})
>e(0;\lambda_{\rm sp})\).
Thus, the bright branch first appears above the
normal energy. Coexistence requires the bright minimum to become
degenerate with the normal minimum:
\(e'(n_{\rm coex})=0\) and \(e(n_{\rm coex})=e(0)\),
with \(n_{\rm coex}>0\).
Eliminating \(\lambda\) gives
\begin{equation}
  \chi n_{\rm coex}^2(\chi n_{\rm coex}+2\omega_c)
  =2\omega_c\omega_0.
  \label{eq:SM_coexistence_unique_condition}
\end{equation}
The left-hand side increases strictly from zero to infinity.
Defining \(r=\chi\omega_0/\omega_c^2>0\) and
\(\Theta=27r/8-1>-1\), its unique positive root is
\begin{equation}
  n_{\rm coex}=\frac{2\omega_c}{3\chi}
  \begin{cases}
    2\cos\bigl[\tfrac13\arccos\Theta\bigr]-1,
    & \Theta\leq1,\\[4pt]
    2\cosh\bigl[\tfrac13\operatorname{arcosh}\Theta\bigr]-1,
    & \Theta>1.
  \end{cases}
  \label{eq:SM_coexistence_density_explicit}
\end{equation}
The corresponding coupling is
\begin{equation}
  \lambda_{\rm coex}^2
  =\frac{(\omega_c+\chi n_{\rm coex})
  (2\omega_c+\chi n_{\rm coex})^2}{4\omega_c}.
  \label{eq:SM_coexistence_coupling_compact}
\end{equation}

Since the energy rises from \(n=0\) to the barrier at \(n_-\),
\(e(n_-)>e(0)\).
Coexistence therefore selects the bright minimum,
\(n_{\rm coex}=n_+(\lambda_{\rm coex})\), with
\(\lambda_{\rm coex}>\lambda_{\rm sp}\).
Along this branch, stationarity removes the chain-rule term
proportional to \(dn_+/d\lambda\), giving
\begin{equation}
  \frac{d}{d\lambda}e[n_+(\lambda);\lambda]
  =-\frac{2\lambda n_+^2}{R(n_+)}<0.
  \label{eq:SM_bright_energy_monotone}
\end{equation}
Because the normal energy remains \(-\omega_0/2\),
this strict decrease establishes the uniqueness of the
coexistence crossing found above.
The bright minimum is metastable for
\(\lambda_{\rm sp}<\lambda<\lambda_{\rm coex}\)
and becomes the global minimum above coexistence.
The normal minimum remains locally stable and is
metastable above coexistence.

For pure thermodynamic phases, the ground-state photon
density is
\begin{equation}
  n_{\rm gs}(\lambda)=
  \begin{cases}
    0, & \lambda<\lambda_{\rm coex},\\
    0\ \text{or}\ n_{\rm coex},
       & \lambda=\lambda_{\rm coex},\\
    n_+(\lambda), & \lambda>\lambda_{\rm coex}.
  \end{cases}
  \label{eq:SM_ground_state_density}
\end{equation}
At coexistence, statistical mixtures of the normal and bright
phases also minimize the energy density and can have any mean
photon density between \(0\) and \(n_{\rm coex}\).
The exact thermodynamic ground-state energy density is
\begin{equation}
  e_\infty(\lambda)=
  \begin{cases}
    -\omega_0/2, & \lambda\leq\lambda_{\rm coex},\\
    e[n_+(\lambda);\lambda],
       & \lambda>\lambda_{\rm coex}.
  \end{cases}
  \label{eq:SM_piecewise_thermodynamic_energy}
\end{equation}
It is continuous at coexistence, while its derivative with
respect to \(\lambda\) jumps from zero to the negative value
in Eq.~\eqref{eq:SM_bright_energy_monotone}, evaluated at
\(\lambda_{\rm coex}\).

At coexistence, define the barrier density
\(n_b=n_-(\lambda_{\rm coex})\).
For \(\omega_c=\omega_0=\chi=1\), \(n_b\simeq0.209756\),
and the barrier per emitter is
\begin{equation}
  \Delta e_b
  =e(n_b;\lambda_{\rm coex})-e(0;\lambda_{\rm coex})
  \simeq0.0890313.
  \label{eq:SM_barrier_height}
\end{equation}
The total CS energy barrier is therefore
\(N\Delta e_b+O(1)\).
Figure~\ref{fig:SM_energy_landscapes} illustrates the five
regimes in density and real-quadrature coordinates:
bright wells appear and become globally stable,
while the normal minimum persists.

\subsection{Dimensionless phase boundaries and weak-Kerr limit}
\label{subsec:SM_dimensionless_boundaries}

Introduce the dimensionless variables
\[
r=\frac{\chi\omega_0}{\omega_c^2},\qquad
x=\frac{\chi n}{\omega_c},\qquad
g=\frac{\lambda}{\omega_c}.
\]
The spinodal condition gives
\begin{align}
  x_{\rm sp}&=\frac{1}{2}\left(\sqrt{1+(4r)^{2/3}}-1\right),\\
 g_{\rm sp}(r)&=(1+x_{\rm sp})^{3/2},
\label{eq:SM_dimensionless_spinodal}
\end{align}
whereas coexistence is determined by
\begin{align}
  x_{\rm coex}^2(x_{\rm coex}+2)&=2r, \nonumber\\
  g_{\rm coex}(r)&=\frac{1}{2}(2+x_{\rm coex})\sqrt{1+x_{\rm coex}}.
  \label{eq:SM_dimensionless_coexistence}
\end{align}
For \(r>0\), the upper-branch selection proved above and \(g_{\rm sp}=(1+x_{\rm sp})^{3/2}>1\) imply
\begin{equation}
  0<x_{\rm sp}<x_{\rm coex},\qquad
  1<g_{\rm sp}(r)<g_{\rm coex}(r).
  \label{eq:SM_scale_ordering}
\end{equation}
Thus, the bright branch appears only beyond the bare collapse coupling. These curves generate Fig.~\ref{fig:main}(a), with \(g_{\rm sp}<g<g_{\rm coex}\) the region of metastable bright minima. On the singular bare line \(r=0\), the model is normal and confined for \(g<1\), reaches spectral collapse at \(g=1\), and is unbounded from below for \(g>1\).

At fixed \(\omega_c,\omega_0\), the weak Kerr limit \(\chi\to0^+\) gives
\begin{align}
  n_{\rm sp}&\sim \frac{\omega_0^{2/3}}{2^{2/3}(\chi\omega_c)^{1/3}},\quad
 n_{\rm coex}\sim\sqrt{\frac{\omega_0}{\chi}},\\
  \lambda_{\rm sp}-\omega_c &\sim\frac{3\omega_c r^{2/3}}{2^{5/3}}, \quad
  \lambda_{\rm coex}-\omega_c \sim\omega_c\sqrt r.
\end{align}
Both couplings approach \(\omega_c\) from above while their bright densities diverge. The confined phase ends singularly at bare collapse, rather than approaching a finite-density bare-model phase.

\subsection{Thermodynamic configurations and observable limits}
\label{subsec:SM_state_observables}

The concentration theorem of Sec.~\ref{sec:SM_exact_convergence} places limiting ground-state Husimi measures on the global minimizers of \(h(\alpha,\boldsymbol\Omega)\). Below coexistence, the unique configuration is \(\alpha=0\), \(\boldsymbol\Omega=(0,0,-1)\). Above coexistence, the four symmetry-related configurations, indexed by \(\ell=0,1,2,3\), are
\begin{align}
  \!\!\alpha_\ell&=i^\ell\sqrt{n_+}, \,\,
  \boldsymbol\Omega_\ell= \left(\!\!-\frac{2(-1)^\ell\lambda n_+}{R(n_+)},\,0,\, -\frac{\omega_0}{R(n_+)}\!\right).
  \label{eq:SM_four_exact_minimizers}
\end{align}
At coexistence, these four points with \(n_+=n_{\rm coex}\) and the normal point are simultaneous minimizers; concentration alone does not fix their relative weights.

Away from coexistence, all minimizing configurations share one photon density \(n_*\). The limiting energy is then differentiable in the model coefficients. Concavity of \(E_0(N)/N\) and variational bounds on its parameter derivatives give the exact observable limits below. For example, \(\partial_\chi e_\infty=n_*^2/2\) fixes the second photon-number moment, not merely the mean:
\begin{align}
  \frac{\langle\hat n\rangle}{N}  \longrightarrow n_*, \quad &\frac{\langle\hat n^2\rangle}{N^2}\longrightarrow n_*^2, \quad \frac{\langle J_z\rangle}{N}\longrightarrow -\frac{\omega_0}{2R_*}, \nonumber\\
  \frac{\langle(a^2+a^{\dagger2})J_x\rangle}{N^2} &\longrightarrow-\frac{2\lambda n_*^2}{R_*},
  \label{eq:SM_exact_observable_limits}
\end{align}
with \(R_*=\sqrt{\omega_0^2+4\lambda^2n_*^2}\). These relations hold for exact ground states, without assuming a product state or finite-\(N\) nondegeneracy. In particular, \(\frac{\operatorname{Var}(\hat n)}{N^2}\longrightarrow0\),
so fluctuations of the photon density vanish. This  is not asserted for normal--bright phase mixtures at coexistence.

For \(\lambda>\lambda_{\rm coex}\), a symmetry-breaking source removed after \(N\to\infty\) selects one bright configuration, giving
\begin{equation*}
  \frac{\langle a\rangle}{\sqrt N}\longrightarrow i^\ell\sqrt{n_*}, \quad
  \frac{\langle J_x\rangle}{N}\longrightarrow -\frac{(-1)^\ell\lambda n_*}{R_*}.
\end{equation*}
Without such selection, a symmetry-preserving ground-state sequence has vanishing \(\langle a\rangle\) and \(\langle J_x\rangle\), even in the bright phase. 

\section{Symmetry-resolved ED and finite-size scaling}
\label{sec:SM_ED_scaling}

In this Section, we combine analytical predictions with symmetry-resolved exact diagonalization (ED) to examine how the thermodynamic normal--bright transition of Secs.~\ref{sec:SM_exact_thermodynamic_limit} and \ref{sec:SM_branch_selection} emerges as \(N\) increases. At each finite \(N\), we scan \(\lambda\) through coexistence at fixed \(\omega_c,\omega_0,\chi\), retaining the \(1/N\) scaling of the pair coupling and Kerr term.  
Three finite-size signatures are examined through \(N=128\): a sharpening photon-density crossover, convergence of the branch-crossing coupling, and exponential closure of the normal--bright tunneling gap. Local Gaussian fluctuations predict the leading \(1/N\) crossing drift (Sec.~\ref{sec:SM_gaussian_fluctuations}). The exponential gap scaling is established numerically through finite-size fits, not derived analytically.

In the maximal-spin basis \(|n_a;j,m\rangle\), with \(j=N/2\), \(0\leq n_a\leq n_a^{\max}\), and \(m=-j,\ldots,j\), the generalized \(\mathbb Z_4\) symmetry separates the Hamiltonian into sectors \(q=[n_a+2(m+j)]\bmod4\), with \(q=0,1,2,3\). At \(\lambda=0\), the normal ground state \(|0;j,-j\rangle\) belongs to \(q=0\). 
Pair-assisted spin flips change \(n_a\) by \(\pm2\) and \(m\) by \(\pm1\), preserving \(q\) modulo four. Thus, normal--bright mixing is possible only within \(q=0\). Levels in different sectors are symmetry-decoupled and may cross exactly. We therefore distinguish the \(q=0\) avoided gap from the splittings involving the other bright-like symmetry sectors.

\subsection{Photon-density curves and cutoff checks}

Figure~\ref{fig:main}(b) shows $q=0$ photon densities for
$N=8,16,32,64,96,128$ over $1.35\leq\lambda\leq2.45$.
Each curve was recomputed at a larger photon cutoff on the same
coupling grid. We monitor the cutoff drift, 
\(\delta n_c=|\langle\hat n\rangle_{\rm main} -\langle\hat n\rangle_{\rm check}|/N,\)
occupation at the upper photon edge, \(w_{\rm edge} =\sum_{n_a=n_a^{\max}-4}^{n_a^{\max}}P(n_a)\),
and eigenpair residual, \(\mathcal R_{\rm eig}=\|H|\psi\rangle-E|\psi\rangle\|_2\). For the six sizes in ascending order, the main/check cutoffs are $80/100$, $100/120$, $120/160$, $220/260$, $300/340$, and $380/440$, respectively; the corresponding grids contain $151,151,141,107,25,25$ coupling points. Across all curves, $\delta n_c\leq3.45\times10^{-11}$, $w_{\rm edge}\leq1.20\times10^{-14}$, and $\mathcal R_{\rm eig}\leq3.77\times10^{-10}$. 
At the finite-size branch crossing defined below, the lowest \(q=0\) eigenstate is an equal-weight superposition of the low-density normal and high-density bright branches. Its photon density is the average of their respective densities.

\subsection{Finite-size crossings and same-sector gaps}
\label{sec:SM_symmetry_gap}

\begin{figure}[t]
\centering
\includegraphics[width=\columnwidth]{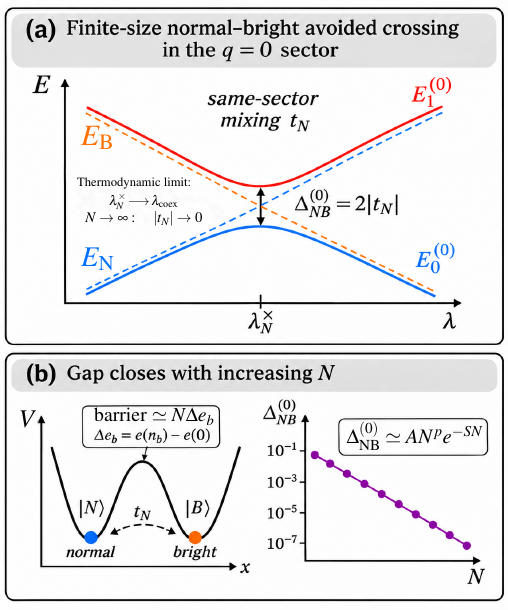}

\caption{\textbf{Normal--bright tunneling and
thermodynamic gap closure.}
(a) Schematic branch energies \(E_{\rm N},E_{\rm B}\)
(dashed) and exact \(q=0\) eigenenergies (solid),
with the common offset
\((E_{\rm N}+E_{\rm B})/2\) subtracted.
At \(\lambda_N^\times\), the branch energies
coincide, while mixing \(t_N\) produces
\(\Delta_{\rm NB}^{(0)}=2|t_N|\).
As \(N\to\infty\),
\(\lambda_N^\times\to\lambda_{\rm coex}\)
and \(|t_N|\to0\).
(b) Schematic collective-coordinate landscape
with a barrier of height \(N\Delta e_b+O(1)\),
where
\(\Delta e_b=e(n_b;\lambda_{\rm coex})
-e(0;\lambda_{\rm coex})\)
and \(n_b=n_-(\lambda_{\rm coex})\).
The semilog plot illustrates the empirical form
\(\Delta_{\rm NB}^{(0)}
\simeq A_\Delta N^p e^{-S_{\rm NB}N}\);
its plotted symbols are illustrative, not ED data.
The barrier height alone does not determine
\(S_{\rm NB}\).}
\label{fig:SM_avoided_crossing_schematic}
\end{figure}

Near coexistence, let $|\psi_0^{(0)}\rangle$ and $|\psi_1^{(0)}\rangle$  be the two lowest exact $q=0$ eigenstates, with energies $E_{0}^{(0)}$ and $E_{1}^{(0)}$. Their spectral projector is $P_2=|\psi_0^{(0)}\rangle\langle\psi_0^{(0)}|+|\psi_1^{(0)}\rangle\langle\psi_1^{(0)}|$. Diagonalizing \(P_2\hat nP_2\) within this doublet defines the low- and high-photon-number branch states \(|{\rm N}\rangle\) and \(|{\rm B}\rangle\), with eigenvalues \(\nu_{\rm N}<\nu_{\rm B}\). In this basis, write
\begin{equation}
  H_{\rm eff}^{(0)}=
  \begin{pmatrix}
  E_{\rm N}&t_N\\
  t_N^*&E_{\rm B}
  \end{pmatrix},
  \label{eq:SM_ED_diabatic_H}
\end{equation}
where \(E_{\rm N}=\langle{\rm N}|H|{\rm N}\rangle\) and \(E_{\rm B}=\langle{\rm B}|H|{\rm B}\rangle\) are the branch energies, while \(t_N=\langle{\rm N}|H|{\rm B}\rangle\) describes their coupling. The finite-size branch-crossing coupling \(\lambda_N^\times\) is defined by \(E_{\rm N}(\lambda_N^\times)=E_{\rm B}(\lambda_N^\times) \equiv E_c\). At \(\lambda=\lambda_N^\times\), \(t_N\) mixes the normal and bright branch states into equal-weight superpositions with exact energies \(E_0^{(0)}=E_c-|t_N|\) and \(E_1^{(0)}=E_c+|t_N|\). Thus, the branch energies cross, but the exact same-sector eigenvalues remain separated (avoid crossing)
whenever \(t_N\neq0\). At \(\lambda_N^\times\), define
\begin{align}
\Delta_{\rm NB}^{(0)}(N)
&\equiv E_1^{(0)}-E_0^{(0)}=2|t_N|,
\label{eq:SM_ED_same_sector_gap}\\
\Delta n_N
&\equiv\frac{\nu_{\rm B}-\nu_{\rm N}}{N}>0.
\label{eq:SM_ED_density_contrast}
\end{align}
Here \(\Delta_{\rm NB}^{(0)}\) measures the splitting of the exact eigenstates, while \(\Delta n_N\) measures the photon-density contrast between the branch states  \(|{\rm N}\rangle\) and \(|{\rm B}\rangle\). 
Figure~\ref{fig:SM_avoided_crossing_schematic}(a) illustrates this distinction.

Table~\ref{tab:SM_symmetry_gap} reports the branch-crossing data and doublet isolation ratios \(\mathcal I_N=\Gamma_N/\Delta_{\rm NB}^{(0)}\), where \(\Gamma_N=E_2^{(0)}-E_1^{(0)}\) is the spacing to the next \(q=0\) level.  The doublets were recomputed at both photon cutoffs using 113-bit symmetric banded shift--invert diagonalization. Maximum cutoff changes in \(\lambda_N^\times\) and \(\Delta n_N\) are \(1.32\times10^{-18}\) and \(6.08\times10^{-17}\), respectively; the largest relative gap change is \(6.58\times10^{-15}\). At \(N=128\), eigenpair residuals are below \(8\times10^{-31}\), and an independent 60-digit solver agrees with the gap to a relative accuracy of \(3.17\times10^{-15}\).

\begin{table}[t]
\caption{Directly computed \(q=0\) data at the branch-crossing coupling \(\lambda_N^\times\), for \(\omega_c=\omega_0=\chi=1\). Here \( \Delta_N = \Delta_{\rm NB}^{(0)}\), and \(\mathcal I_N = (E_2^{(0)} - E_1^{(0)})/\Delta_N\) compares the spacing to the next \(q=0\) level with the splitting within the doublet.}
\label{tab:SM_symmetry_gap}
\centering
\setlength{\tabcolsep}{2.0pt}
\renewcommand{\arraystretch}{1.04}
\begin{ruledtabular}
\begin{tabular}{@{}rcccc@{}}
\(N\) & \(\lambda_N^\times\) & \(\Delta_N\) &
\(\Delta n_N\) & \(\mathcal I_N\)\\
\hline
8   & 1.978244 & \(2.588035\times10^{-1}\)  & 0.764956 & 5.37\\
16  & 1.946466 & \(3.099560\times10^{-2}\)  & 0.784532 & \(4.32\times10^1\)\\
24  & 1.938956 & \(3.004858\times10^{-3}\)  & 0.805485 & \(5.29\times10^2\)\\
32  & 1.935411 & \(2.674301\times10^{-4}\)  & 0.814869 & \(6.65\times10^3\)\\
40  & 1.933332 & \(2.284936\times10^{-5}\)  & 0.820135 & \(8.09\times10^4\)\\
48  & 1.931964 & \(1.905559\times10^{-6}\)  & 0.823522 & \(9.86\times10^5\)\\
56  & 1.930996 & \(1.563756\times10^{-7}\)  & 0.825887 & \(1.21\times10^7\)\\
64  & 1.930274 & \(1.268551\times10^{-8}\)  & 0.827634 & \(1.51\times10^8\)\\
72  & 1.929716 & \(1.020196\times10^{-9}\)  & 0.828976 & \(1.89\times10^9\)\\
80  & 1.929271 & \(8.149410\times10^{-11}\) & 0.830041 & \(2.37\times10^{10}\)\\
88  & 1.928907 & \(6.474680\times10^{-12}\) & 0.830907 & \(3.00\times10^{11}\)\\
96  & 1.928605 & \(5.121343\times10^{-13}\) & 0.831624 & \(3.80\times10^{12}\)\\
128 & 1.927778 & \(1.941664\times10^{-17}\) & 0.833577 & \(1.01\times10^{17}\)
\end{tabular}
\end{ruledtabular}
\end{table}

Two additional tests assess the branch construction beyond the doublet. At \(\lambda_N^\times\), a number bias \(H_h=H-h\hat n\), with \(|h|=\Delta_{\rm NB}^{(0)}/(N\Delta n_N)\), predicts a favored-branch probability \((1+1/\sqrt2)/2\simeq0.853553\) within the two-state model. Full-\(q=0\) diagonalization gives \(0.8491/0.8582\) for the two bias signs at \(N=8\), and \(0.8532/0.8540\) at \(N=16\), with maximum weights outside \(P_2\) of \(3.83\times10^{-3}\) and \(3.41\times10^{-5}\), respectively.

For a four-state check, let \(P_4\) project onto the four lowest \(q=0\) states. We diagonalize \(P_4\hat nP_4\), partition its eigenvectors into low- and high-density groups at \((\nu_{\rm N}+\nu_{\rm B})/2\), and diagonalize \(H\) within each group. Relative to the doublet construction, the crossing and density-contrast changes are \(1.28\times10^{-2}\) and \(1.10\times10^{-1}\) at \(N=8\), but below \(4.4\times10^{-8}\) and \(5.8\times10^{-7}\) for \(N\geq32\). The extrapolated crossing intercept changes by \(5.90\times10^{-6}\). These checks assess the same-sector branch construction. Inter-sector splittings are examined separately below.

\subsection{Inter-sector bright-multiplet splittings}
\label{sec:SM_intersector_splittings}

The remaining three bright-like states belong to the symmetry-decoupled \(q=1,2,3\) sectors. At the finite-size branch crossing \(\lambda_N^\times\), let \(E_0^{(q)}\) be the lowest exact energy in sector \(q\), and \(E_{\rm B}=\langle{\rm B}|H|{\rm B}\rangle\) the high-density branch energy defined above. Their energy offsets are
\[
\delta_q=E_0^{(q)}-E_{\rm B},\qquad
\delta_{\min}=\min_{q=1,2,3}\delta_q.
\]
Since \(E_0^{(0)}=E_{\rm B}-|t_N|\) at the branch crossing, the gap from the \(q=0\) ground state to the lowest state in another symmetry sector of the maximal-spin manifold is
\begin{equation}
  \Delta_p(N) =\min_{q=1,2,3} \left[E_0^{(q)}-E_0^{(0)}\right] =|t_N|+\delta_{\min}.
  \label{eq:SM_intersector_gap}
\end{equation}
Defining \(\eta_N=\max_{q=1,2,3}|\delta_q|/|t_N|\), we obtain
\begin{equation}
\left| \frac{2\Delta_p}{\Delta_{\rm NB}^{(0)}}-1 \right|\leq\eta_N.    
\end{equation}
Table~\ref{tab:SM_intersector_offsets} shows \(\eta_N<1\), hence \(\Delta_p>0\), for all tabulated sizes. The lowest \(q\neq0\) level belongs to \(q=2\), with \(\delta_2<0\). Thus, at these crossings, \(\Delta_p\) is the lowest excitation gap within the maximal-spin manifold, whereas \(\Delta_{\rm NB}^{(0)}\) is the distinct
avoided splitting within \(q=0\).

The decreasing \(\eta_N\) indicates that the bright-multiplet offsets become small relative to the normal--bright coupling. Fitting \(\eta_N\propto N^{p_\eta}e^{-\Delta S N}\) over \(16\leq N\leq48\) gives \(\Delta S=0.330\), or \(0.326\) when the lower fit boundary is raised to \(N=24\). Separate fits \(|\delta_q|\propto N^{p_q}e^{-S_qN}\) give \((S_1,S_2,S_3)=(0.662,0.656,0.663)\), exceeding the fitted normal--bright exponent \(S_{\rm NB}=0.324\). The data are therefore consistent with the bright-multiplet offsets becoming exponentially subleading to \(|t_N|\), implying \(\Delta_p/\Delta_{\rm NB}^{(0)}\to1/2\) if this trend persists asymptotically. This limiting relation is supported by finite-size fits, not established analytically.

\subsection{Finite-size drift and exponential gap closure}

The finite-\(N\) branch crossing differs from thermodynamic coexistence as the normal and bright branches acquire different quantum energy corrections. Expanding the Hamiltonian to quadratic (Gaussian) order about each stationary minimum gives the branch energies
\begin{equation} 
  E_\beta(N,\lambda) = Ne_\beta(\lambda) + \gamma_\beta(\lambda) +  O(N^{-1}),
\label{eq:SM_ED_branch_energy_expansion}
\end{equation}
for \(\beta\in\{{\rm N},{\rm B}\}\), where \(e_\beta\) is the thermodynamic energy density and \(\gamma_\beta\) the \(O(1)\) fluctuation correction. For the \(\hat n^2\) Kerr convention, \(\gamma_{\rm N}=0\), while the bright-well Gaussian modes give
\begin{equation}
  \gamma_{\rm B} = \frac{\varepsilon_++\varepsilon_--A-R}{2},
  \label{eq:SM_gamma_bright}
\end{equation}
where 
\(\varepsilon_\pm\) are the coupled cavity--spin fluctuation frequencies derived in Sec.~\ref{sec:SM_gaussian_fluctuations}, evaluated on the stationary bright branch.

\begin{table}[t]
\caption{Bright-multiplet offsets at
\(\lambda_N^\times\) for \(\omega_c=\omega_0=\chi=1\).
All \(\delta_q<0\) over the displayed range, so
absolute values are shown. The lowest \(q\neq0\)
state belongs to \(q=2\).}
\label{tab:SM_intersector_offsets}
\centering
\scriptsize
\setlength{\tabcolsep}{1pt}
\renewcommand{\arraystretch}{1.05}
\begin{ruledtabular}
\begin{tabular}{@{}rcccc@{}}
\(N\) & \(|\delta_1|\) & \(|\delta_2|\) &
\(|\delta_3|\) & \(\eta_N\)\\
\hline
8  & \(3.43030\times10^{-2}\) & \(5.38791\times10^{-2}\) & \(2.71571\times10^{-2}\) & \(4.16370\times10^{-1}\)\\
16 & \(7.16284\times10^{-4}\) & \(1.09914\times10^{-3}\) & \(5.93905\times10^{-4}\) & \(7.09222\times10^{-2}\)\\
24 & \(7.48162\times10^{-6}\) & \(1.14776\times10^{-5}\) & \(6.30309\times10^{-6}\) & \(7.63936\times10^{-3}\)\\
32 & \(6.19996\times10^{-8}\) & \(9.67126\times10^{-8}\) & \(5.25343\times10^{-8}\) & \(7.23274\times10^{-4}\)\\
40 & \(4.62362\times10^{-10}\) & \(7.37754\times10^{-10}\) & \(3.92584\times10^{-10}\) & \(6.45754\times10^{-5}\)\\
48 & \(3.24187\times10^{-12}\) & \(5.30858\times10^{-12}\) & \(2.75298\times10^{-12}\) & \(5.57168\times10^{-6}\)
\end{tabular}
\end{ruledtabular}
\end{table}

At \(\lambda_N^\times\), \(E_{\rm N}=E_{\rm B}\). Expanding this equality about \(\lambda_{\rm coex}\), where the leading branch energies coincide, gives
\begin{equation}
  \lambda_N^\times = \lambda_{\rm coex}+\frac{c_1}{N}+O(N^{-2}), \quad
  c_1=\frac{\gamma_{\rm B}-\gamma_{\rm N}}{\ell_\lambda},
  \label{eq:SM_analytic_c1}
\end{equation}
where \(\ell_\lambda =-\partial_\lambda(e_{\rm B}-e_{\rm N}) |_{\lambda_{\rm coex}} = 2\lambda_{\rm coex}n_{\rm coex}^2/ R_{\rm coex}>0\).

For \( \omega_c = \omega_0 = \chi = 1\), the Gaussian calculation gives \(\gamma_{\rm B}=0.249675\), \( \ell_\lambda = 0.801781\), and \(c_1 = 0.311401\). 
Holding \(\lambda_{\rm coex}\) fixed, an ED fit of the unrounded crossings to \(\lambda_N^\times =\lambda_{\rm coex}+c_1/N+c_2/N^2\) for \(N\geq40\) gives \(c_1^{\rm ED}=0.311062\), differing from the Gaussian prediction by only \(0.109\%\).

In contrast, the avoided gap within \(q=0\) arises from quantum mixing between the normal and bright branches. The ED results are well described by
\begin{equation}
  \Delta_{\rm NB}^{(0)}(N) \simeq A_\Delta N^p e^{-S_{\rm NB}N}.
  \label{eq:SM_ED_gap_fit_form}
\end{equation}
An unweighted least-squares fit of $\log\Delta_{\mathrm{NB}}^{(0)}(N)$ to
$\log A_\Delta+p\log N-S_{\mathrm{NB}}N$ over $16\leq N\leq96$ yields
\begin{equation}
  A_\Delta=1.02366,\,\, p=0.611063,\,\, S_{\rm NB}=0.324130.
  \label{eq:SM_ED_gap_fit}
\end{equation}
Raising the lower fit boundary to \(N=24\)--\(48\) gives \(S_{\rm NB}=0.3231\)--\(0.3236\), indicating a stable fitted exponential rate. The fit excludes \(N=128\) and predicts \(\Delta_{\rm NB}^{(0)} = 1.90319\times10^{-17}\), compared with the directly computed \(1.94166\times10^{-17}\), a \(1.98\%\) deviation relative to the directly computed gap. The data support exponential suppression, but do not determine an analytical tunneling exponent; the algebraic prefactor is also less well constrained. Figure~\ref{fig:SM_avoided_crossing_schematic}(b) illustrates the coherent-state barrier and this gap suppression without identifying the barrier height with the tunneling exponent.

Thus, the \(1/N\) crossing drift is set by local Gaussian energy corrections, whereas the exponentially closing same-sector gap measures quantum mixing between the normal and bright configurations. Together with the sharpening photon-density crossover, these provide consistent finite-size signatures of first-order coexistence.




\section{Local Gaussian fluctuations and finite-$N$ validation}
\label{sec:SM_gaussian_fluctuations}

The preceding sections establish the normal--bright
phase structure and the exponentially closing
finite-$N$ tunneling gap. Here we examine how the cavity field and collective spin fluctuate around each stable configuration. A quadratic (Gaussian) expansion yields the local
collective modes and their finite-size energy
corrections, while the Gaussian covariances
characterize spin squeezing and cavity--spin
entanglement. We test the predicted bright-state
spin fluctuations using localized finite-$N$
states constructed from the four $\mathbb Z_4$
symmetry sectors. Numerical examples use $\omega_c=\omega_0=\chi=1$.

\subsection{Local quadratic Hamiltonians}
\label{subsec:SM_local_quadratic}

To study local quantum excitations, we expand \(H\) about each stationary field--spin configuration as \( N\to\infty\), keeping \( \omega_c, \omega_0, \chi, \lambda\) fixed. 
On the normal branch, the cavity amplitude vanishes and the collective spin points along \(-z\), i.e., \(\alpha=0\) and \(\boldsymbol\Omega=(0,0,-1)\). We set \(c=a\) and introduce a Holstein--Primakoff boson \(d\) to describe spin excitations about this configuration:
\begin{equation}
  J_z=d^\dagger d-\frac N2,\quad
  J_x=\frac{\sqrt N}{2}(d+d^\dagger)+O(N^{-1/2}).
\label{eq:SM_normal_HP}
\end{equation}
Thus, \(d^\dagger d\) counts deviations from complete downward polarization. For \(O(1)\) fluctuation occupations, the pair interaction first contributes at cubic order \(O(N^{-1/2})\), while the Kerr term is \(O(N^{-1})\). Neither enters the leading quadratic Hamiltonian,
\begin{equation}
  H_{\rm N}^{(2)} = \omega_c c^\dagger c + \omega_0 d^\dagger d.
  \label{eq:SM_normal_gaussian}
\end{equation}
Here the classical normal-state energy \(Ne(0)=-\omega_0N/2\) has been subtracted. Thus the normal modes retain frequencies \(\omega_c,\omega_0\), with no local softening at coexistence. Their ground state has transverse spin variances
\begin{equation}
  \frac4N\operatorname{Var}(J_x) = \frac4N\operatorname{Var}(J_y) = 1+O(N^{-1}).
 \label{eq:SM_normal_unsqueezed}
\end{equation}

For the bright branch, choose the real minimum \(\alpha=\sqrt n\), with \(n=n_+(\lambda)\), and displace the cavity field as \(a=\sqrt{Nn}+c\). The leading spin-dependent Hamiltonian is \(H_{\rm spin}=\omega_0J_z+2\lambda nJ_x =\boldsymbol B_{\rm eff}\cdot\mathbf J\), with \(\boldsymbol B_{\rm eff}=(2\lambda n,0,\omega_0)\). Its minimum occurs when the collective spin points antiparallel to \(\boldsymbol B_{\rm eff}\). Defining \(R=|\boldsymbol B_{\rm eff}|=\sqrt{\omega_0^2+4\lambda^2n^2}\), the spin tilt from \(-z\) toward \(-x\) satisfies
\begin{equation}
  \sin\vartheta=\frac{2\lambda n}{R}, \qquad
  \cos\vartheta=\frac{\omega_0}{R}.
  \label{eq:SM_bright_angle}
\end{equation}
The stationary spin orientation therefore gives
\begin{equation}
  \frac{\langle J_x\rangle}{N} = -\frac{\lambda n}{R},\qquad
  \frac{\langle J_z\rangle}{N} =-\frac{\omega_0}{2R},\qquad
  \langle J_y\rangle=0.
  \label{eq:SM_bright_spin_mean}
\end{equation}
To expand about this tilted configuration, rotate
the spin operators into a local frame:
\begin{align}
  J_x&=\cos\vartheta\,\widetilde J_x +\sin\vartheta\,\widetilde J_z,\quad J_y=\widetilde J_y \nonumber\\
  J_z&=-\sin\vartheta\,\widetilde J_x +\cos\vartheta\,\widetilde J_z.
\label{eq:SM_bright_rotation}
\end{align}
In this frame, \(H_{\rm spin}=R\widetilde J_z\). The reference spin again points along \(-z\) in this frame, allowing the Holstein--Primakoff expansion. The other three \(\mathbb Z_4\)-related bright minima have identical local fluctuation spectra.


In the rotated spin frame, we use \(\widetilde J_z=d^\dagger d-N/2\) and \(\widetilde J_x=\sqrt N(d+d^\dagger)/2 + O(N^{-1/2})\). Expanding the Hamiltonian to quadratic order in the cavity and spin fluctuations \(c,d\), and subtracting the classical energy \(Ne(n)\), gives
\begin{align}
  H_{\rm B}^{(2)} = {}&A c^\dagger c +\frac{\chi n}{2}(c+c^\dagger)^2 -\frac A2(c^2+c^{\dagger2})\nonumber\\
  &+ R d^\dagger d +\mathcal G(c+c^\dagger)(d+d^\dagger),
\label{eq:SM_bright_quadratic_boson}
\end{align}
where \(A=\omega_c+\chi n=2\lambda^2n/R\) follows from stationarity, and \(\mathcal G=\lambda\sqrt n\,\omega_0/R\) is the local cavity--spin coupling.
In terms of the canonical quadratures
\[
x_a=\frac{c+c^\dagger}{\sqrt2},\,
p_a=\frac{c-c^\dagger}{i\sqrt2},\,
x_s=\frac{d+d^\dagger}{\sqrt2},\,
p_s=\frac{d-d^\dagger}{i\sqrt2},
\]
we can also rewrite  \(H_{\rm B}^{(2)}\) compactly as
\begin{equation}
  H_{\rm B}^{(2)} = \frac12\mathbf p^T\mathsf P\mathbf p +\frac12\mathbf x^T\mathsf K\mathbf x -\frac{A+R}{2},
  \label{eq:SM_bright_quadratic}
\end{equation}
where \(\mathbf x=(x_a,x_s)^T\), \(\mathbf p=(p_a,p_s)^T\), and
\begin{equation}
  \mathsf P=
  \begin{pmatrix}2A&0\\0&R\end{pmatrix}, \qquad
  \mathsf K=
  \begin{pmatrix}
  2\chi n&2\mathcal G\\
  2\mathcal G&R
  \end{pmatrix}.
  \label{eq:SM_full_Hessian}
\end{equation}
Here \(\mathsf P\) determines the momentum-dependent
terms, while \(\mathsf K\) describes the restoring
forces and coupling between cavity and spin fluctuations.
Since \(\mathsf P\) is positive definite, local stability
requires \(\mathsf K\) to be positive definite.
The eigenvalues of
\(\mathsf P^{1/2}\mathsf K\mathsf P^{1/2}\)
give the squared collective-mode frequencies, while
the corresponding Gaussian covariances determine
spin squeezing and cavity--spin entanglement.
We evaluate these quantities below.

\subsection{Collective modes and spinodal softening}
\label{subsec:SM_local_modes}

The Heisenberg equations for Eq.~\eqref{eq:SM_bright_quadratic} give \(\dot{\mathbf x}=\mathsf P\mathbf p\) and \(\dot{\mathbf p}=-\mathsf K\mathbf x\), hence \(\ddot{\mathbf x}=-\mathsf P\mathsf K\mathbf x\). Since \(\mathsf P\) is positive definite, \(\mathsf P\mathsf K\) is similar to the symmetric matrix \(\mathsf D=\mathsf P^{1/2}\mathsf K\mathsf P^{1/2}\). Its eigenvalues give the squared collective-mode frequencies,
\begin{equation}
  \!\!\varepsilon_\pm^2 \!=\!\frac12\!\left[ R^2 \!+\! 4A\chi n \pm \!\sqrt{(R^2 \!-\! 4A\chi n)^2 \!+\! 16A^2\omega_0^2} \right].
\label{eq:SM_bright_frequencies}
\end{equation}

Let \(\mathsf O\) be the orthogonal matrix satisfying \(\mathsf O^T\mathsf D\mathsf O = \operatorname{diag}(\varepsilon_+^2,\varepsilon_-^2)\). The canonical normal-mode coordinates are \(\mathbf Q=\mathsf O^T\mathsf P^{-1/2}\mathbf x\) and \( \boldsymbol\Pi = \mathsf O^T\mathsf P^{1/2}\mathbf p\). Defining \( b_\sigma = [\sqrt{\varepsilon_\sigma}Q_\sigma+ i\Pi_\sigma/\sqrt{\varepsilon_\sigma}]/\sqrt2\), we obtain
\begin{equation} 
H_{\rm B}^{(2)} = \sum_{\sigma=\pm}\varepsilon_\sigma \Big(b_\sigma^\dagger b_\sigma+\frac12\Big) -\frac{A+R}{2},
\end{equation}
where \(b_\pm\) annihilate the two collective modes. Their vacuum gives the \(O(N^0)\) bright-branch energy correction
\begin{equation}
  \gamma_{\rm B} = \langle H_{\rm B}^{(2)}\rangle =\frac{\varepsilon_++\varepsilon_--A-R}{2},
\end{equation}
used in Eq.~\eqref{eq:SM_gamma_bright} to determine the  finite-\(N\) crossing shift.

The stability of the stationary bright configurations is determined by the curvature of the thermodynamic energy density. We find
\begin{equation}
  e''(n)=\chi-\frac{2\lambda^2\omega_0^2}{R^3}, \qquad
  \det\mathsf K=2nR\,e''(n).
  \label{eq:SM_Hessian_curvature_identity}
\end{equation}
Since \(\mathsf P\) is positive definite and \(\mathsf K_{11}=2\chi n>0\), local stability requires \(\det\mathsf K>0\), or equivalently \(e''(n)>0\). Thus the upper-density stationary solution \(n_+\) is a stable bright minimum, whereas the lower-density solution \(n_-\), with \(e''(n_-)<0\), has one unstable direction. At the spinodal, the two solutions merge and \(e''(n_{\rm sp})=0\).

The connection to the excitation spectrum follows from
\begin{equation}
  \varepsilon_+^2\varepsilon_-^2 = \det\mathsf D=4AnR^2e''(n).
  \label{eq:SM_frequency_curvature_identity}
\end{equation}
So far, we have written \(e(n)\), suppressing its dependence on \(\lambda\), with primes denoting derivatives with respect to \(n\) at fixed \(\lambda\). We now write \(e(n;\lambda)\) explicitly because both quantities vary near the spinodal.

At the spinodal, the two stationary bright branches merge, satisfying \(e'(n_{\rm sp};\lambda_{\rm sp}) = e''(n_{\rm sp};\lambda_{\rm sp})=0\). Writing \(\delta n=n_+(\lambda)-n_{\rm sp}\) and \(\delta\lambda=\lambda-\lambda_{\rm sp}>0\), the stationary-point equation gives, to leading order,
\begin{equation}
  0=e'(n_+;\lambda)\simeq (\partial_\lambda e')_{\rm sp}\delta\lambda + \frac12e'''_{\rm sp}(\delta n)^2.
\end{equation}
Since \((\partial_\lambda e')_{\rm sp}<0\) and \(e'''_{\rm sp}>0\), we obtain \(\delta n\propto(\delta\lambda)^{1/2}\) and consequently \(e''(n_+;\lambda)\simeq e'''_{\rm sp}\delta n \propto(\delta\lambda)^{1/2}\).

At the spinodal, Eq.~\eqref{eq:SM_frequency_curvature_identity} gives \(\varepsilon_+^2\varepsilon_-^2=0\), while their sum \(\varepsilon_+^2+\varepsilon_-^2 =R^2+4A\chi n\) remains positive. Thus \(\varepsilon_+\) stays finite and \(\varepsilon_-\to0\). Since \(\varepsilon_-^2\propto e''(n_+;\lambda)\), we obtain
\begin{equation}
  \varepsilon_-\propto (\lambda-\lambda_{\rm sp})^{1/4}.
  \label{eq:SM_quarter_power}
\end{equation}
The prefactor \( \left( \frac{4A_{\rm sp}n_{\rm sp}R_{\rm sp}^2} {R_{\rm sp}^2+4A_{\rm sp}\chi n_{\rm sp}} e'''_{\rm sp}C_n \right)^{1/2} = 2.1422\) for \(\omega_c=\omega_0=\chi=1\), where \(C_n= \left(-\frac{2(\partial_\lambda e')_{\rm sp}} {e'''_{\rm sp}}\right)^{1/2}\). This local softening at the spinodal is distinct from the exponentially closing normal--bright tunneling gap at coexistence.

\subsection{Spin squeezing and cavity--spin entanglement}
\label{subsec:SM_gaussian_correlations}

Although the Gaussian ground state \(|0_{\rm G}\rangle\) satisfies \(b_\pm|0_{\rm G}\rangle=0\), the collective modes mix the cavity and spin fluctuations \(c,d\) and their adjoints. Thus, their vacuum can exhibit correlations between the original cavity and spin degrees of freedom. To quantify these correlations and the individual fluctuations, we use the coordinate and momentum covariance matrices. Their diagonal elements give the respective variances, while the off-diagonal elements describe cavity--spin correlations. For the zero-mean Gaussian ground state,
\begin{align}
  \boldsymbol\Gamma_x &=\begin{pmatrix}
\langle x_a^2\rangle&
\langle x_ax_s\rangle\\
\langle x_sx_a\rangle&
\langle x_s^2\rangle
\end{pmatrix}= \frac12\mathsf P^{1/2}\mathsf D^{-1/2}\mathsf P^{1/2},\\
  \boldsymbol\Gamma_p &= 
\begin{pmatrix}
\langle p_a^2\rangle&
\langle p_ap_s\rangle\\
\langle p_sp_a\rangle&
\langle p_s^2\rangle
\end{pmatrix}
= \frac12\mathsf P^{-1/2}\mathsf D^{1/2}\mathsf P^{-1/2}.
  \label{eq:SM_gaussian_covariances}
\end{align}
\((\boldsymbol\Gamma_x)_{ij} = \langle\{x_i,x_j\}\rangle/2\), and similarly for momenta. Since the quadratic Hamiltonian has no mixed \(xp\) terms, the symmetrized coordinate--momentum covariances vanish.

Recall the Holstein–Primakoff relations \( \widetilde J_x\simeq \frac{\sqrt N}{2}(d+d^\dagger)\), and \( \widetilde J_y = J_y\simeq\frac{\sqrt N}{2i}(d-d^\dagger)\). Using the spin quadratures, \(x_s\) and \(p_s\),
these become \( \widetilde J_x\simeq\sqrt{\frac N2}\,x_s\), and \( J_y\simeq\sqrt{\frac N2}\,p_s\).
Consequently, \(\operatorname{Var}(\widetilde J_x)=\frac N2\langle x_s^2\rangle\), and \(\operatorname{Var}(J_y) =\frac N2\langle p_s^2\rangle\). Since the spin quadratures correspond to the \(ss\) entries of the covariance matrices,
\begin{align}
  \xi_{\widetilde x}^2 &\equiv\frac4N\operatorname{Var}(\widetilde J_x) =2(\boldsymbol\Gamma_x)_{ss},\,\,
  \xi_y^2 \equiv\frac4N\operatorname{Var}(J_y) =2(\boldsymbol\Gamma_p)_{ss}.\nonumber
\end{align}
This Kitagawa--Ueda normalization compares each transverse variance with the coherent-spin value \(N/4\)~\cite{Kitagawa_PRA_1993}. With this normalization, an unsqueezed coherent spin has \(\xi_{\widetilde x}^2=\xi_y^2=1\). It differs from Wineland squeezing, which also includes the mean-spin length~\cite{Wineland_PRA_1992}. Writing \(u=\varepsilon_+\varepsilon_-\), \(v=\varepsilon_++\varepsilon_-\), and \(\mathcal C=4A\chi n\),
\begin{equation}
  \xi_{\widetilde x}^2=\frac{R(\mathcal C+u)}{uv},\qquad
  \xi_y^2=\frac{R^2+u}{Rv}.
\label{eq:SM_spin_squeezing_closed}
\end{equation}
For \(\mathcal G\neq0\), \(\varepsilon_-<R<\varepsilon_+\), so
\begin{equation}
  \xi_y^2-1= \frac{(R-\varepsilon_+)(R-\varepsilon_-)}{Rv}<0.
  \label{eq:SM_generic_bright_squeezing}
\end{equation}
Thus every stable coupled bright well exhibits Gaussian variance squeezing along $J_y$. Moreover, $\xi_{\widetilde x}^2\xi_y^2 = 1 + 4A^2\omega_0^2/(uv^2)>1$, implying anti-squeezing along $\widetilde J_x$. Near the spinodal,
\begin{equation}
  \xi_{\widetilde x}^2\propto(\lambda-\lambda_{\rm sp})^{-1/4},\,\,\,
  \xi_y^2\longrightarrow \frac{R_{\rm sp}}{\sqrt{R_{\rm sp}^2+4A_{\rm sp}\chi n_{\rm sp}}}.
  \label{eq:SM_spinodal_spin_fluctuations}
\end{equation}
For unit parameters, the limiting \(\xi_y^2\) is \(0.75216\). These statements take \(N\to\infty\) first at fixed positive \(\lambda-\lambda_{\rm sp}\); the diverging fluctuations make the Gaussian expansion nonuniform in a shrinking neighborhood of the spinodal. At coexistence, both local modes remain finite:
\[
\varepsilon_- \simeq2.06997,\,
\varepsilon_+ \simeq3.65165,\,
\xi_{\widetilde x}^2 \simeq 1.07426,\,
\xi_y^2 \simeq0.981775.
\]

The local Gaussian ground state is a pure two-mode state of the cavity and spin fluctuations. Tracing out the cavity leaves a reduced spin covariance matrix
\begin{equation}
  \boldsymbol\Gamma_s =\frac12
\begin{pmatrix}
\xi_{\widetilde x}^2&0\\
0&\xi_y^2
  \end{pmatrix}.
\end{equation}
Its symplectic eigenvalue \(\nu_s=\sqrt{\det\boldsymbol\Gamma_s} = \sqrt{\xi_{\widetilde x}^2\xi_y^2}/2\) determines the cavity--spin logarithmic negativity (using natural logarithms):
\begin{equation}
  E_{\mathcal N} =\operatorname{arcosh}(2\nu_s) =\operatorname{arcosh} \sqrt{\xi_{\widetilde x}^2\xi_y^2} \simeq0.23177.
  \label{eq:SM_coexistence_negativity}
\end{equation}
Thus, the reduced spin state is mixed (\(\nu_s>1/2\)), demonstrating entanglement with the cavity. These results characterize a single bright well, not a symmetry-preserving superposition of wells.

Figure~\ref{fig:SM_gaussian_summary}(a)--(c) summarizes the mode spectrum, squeezing, and spinodal scaling.

\subsection{Finite-\texorpdfstring{$N$}{N} localized-multiplet validation}
\label{sec:SM_localized_bright_validation}

\begin{figure}[t]
\centering
\includegraphics[width=\columnwidth,height=1.75in]
{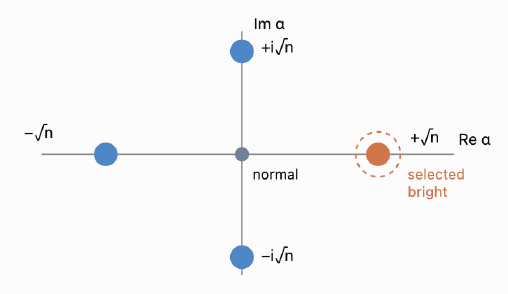}
\caption{\textbf{Localized bright-state construction.}
Four symmetry-related thermodynamic bright minima
in the intensive field-amplitude plane
\(\alpha=\langle a\rangle/\sqrt N\).
The origin represents the normal configuration.
Finite-\(N\) symmetry eigenstates have
\(\langle a\rangle=0\); their phase-matched
superposition selects the highlighted real
bright configuration.}
\label{fig:SM_bright_localization}
\end{figure}

The preceding Gaussian theory describes fluctuations around a single bright minimum. The thermodynamic energy landscape, however, has four symmetry-related minima, \(\alpha_\ell=i^\ell\sqrt{n_+}\) (\(\ell=0,1,2,3\)), as shown in Fig.~\ref{fig:SM_bright_localization}. At finite \(N\), energy eigenstates can be chosen to have definite \(\mathbb Z_4\) symmetry quantum number \(q\). Since \(a\) changes the symmetry sector, each such state satisfies \(\langle a\rangle=0\), even when its photon occupation \(\langle a^\dagger a\rangle\sim Nn_+\), i.e., \(O(N)\). Thus, a symmetry eigenstate cannot directly represent an individual bright minimum, for which \(\langle a\rangle/\sqrt N=\sqrt{n_+}\) in the selected real configuration.

To test the Gaussian predictions against finite-\(N\) ED, we construct a state localized near this real minimum. Let \(|B_q\rangle\) denote the lowest bright-like state in symmetry sector \(q=0,1,2,3\). Choosing their relative phases such that \(\langle B_{q-1}|a|B_q\rangle>0\) for \(q=1,2,3\), we form
\begin{equation}
  |B_{\ell=0}^{\rm loc}\rangle =\frac12\sum_{q=0}^3|B_q\rangle.
\label{eq:SM_localized_bright_state}
\end{equation}
This superposition need not be an energy eigenstate. Its purpose is to select one bright configuration and provide a finite-\(N\) counterpart of the local Gaussian state.

Table~\ref{tab:SM_gaussian_ED_comparison}
summarizes the two descriptions. Gaussian theory
predicts the mean field, spin orientation, and
quantum fluctuations around the selected minimum,
while localized ED states allow these predictions
to be tested at finite \(N\).

\begin{table}[t]
\caption{\textbf{Gaussian theory versus finite-\(N\) ED.}
Comparison for the same selected bright configuration.}
\label{tab:SM_gaussian_ED_comparison}
\begin{ruledtabular}
\begin{tabular}{
@{}p{0.48\columnwidth}
@{\hspace{0.03\columnwidth}}
p{0.46\columnwidth}@{}
}
\textbf{Gaussian theory (\(N\to\infty\))}
& \textbf{Finite-\(N\) ED}\\
\hline
Expansion around one bright minimum
& Superposition of four \(\mathbb Z_4\)-sector states\\
Quadratic cavity--spin fluctuations
& Full finite-\(N\) Hamiltonian\\
Predicts local transverse spin fluctuations
& Calculates the corresponding variances\\
Provides thermodynamic limiting values
& Tests convergence as \(N\) increases
\end{tabular}
\end{ruledtabular}
\end{table}

\begin{figure*}[t]
\centering
\includegraphics[width=\textwidth,height=10cm]{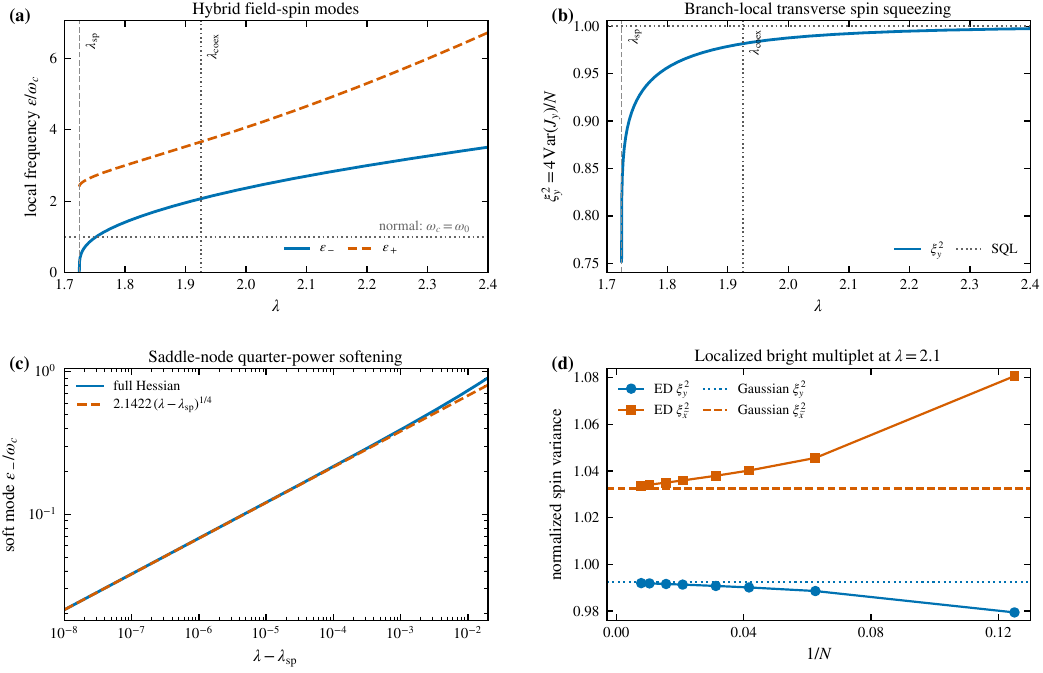}\\
\vspace{-2ex}
\caption{\textbf{Local Gaussian fluctuations and finite-\(N\) validation.}
All panels use \(\omega_c=\omega_0=\chi=1\).
(a) Coupled cavity--spin frequencies on the bright-minimum branch. The normal frequencies remain \(\omega_c=\omega_0=1\).
(b) Transverse bright-well variance \(\xi_y^2\), normalized to the
coherent-spin value (SQL).
(c) Quarter-power softening at the saddle-node spinodal.
(d) Variances of localized bright wave packets at \(\lambda=2.1\),
for \(N=8,16,24,32,48,64,96,128\), compared with the Gaussian limits.
The finite-\(N\) \(\widetilde x\) axis is transverse to the measured
mean spin. All Gaussian curves describe individual wells. They do
not represent the normal--bright tunneling gap.}
\label{fig:SM_gaussian_summary}
\end{figure*}

At \(\lambda=2.1\), Fig.~\ref{fig:SM_gaussian_summary}(d) shows that the transverse spin variances of the localized finite-\(N\) states approach their thermodynamic Gaussian predictions as \(N\) increases. The \(\widetilde x\) axis is defined relative to the finite-\(N\) mean-spin direction. Repeating the calculations at larger photon cutoffs gives a maximum observable change of \(7.9\times10^{-13}\), confirming numerical convergence.



\section{Passive one-photon loss and mean-field stability}
\label{sec:SM_passive}

Energetic stability and dissipative dynamical stability are distinct. A bright minimum of the closed Hamiltonian need not be an attractor under dissipation. We therefore ask when our model supports locally attracting bright fixed points under passive loss and dephasing.

For the bare two-photon Dicke model ($\chi=0$), previous work established that ordinary one-photon loss does not remove the pair-amplification instability, whereas two-photon loss can stabilize superradiant steady states~\cite{Shah2025}. These results demonstrate the role of nonlinear dissipation in the bare model but do not determine the effect of Hamiltonian Kerr confinement.

\subsection{Self-consistent passive fixed points}
\label{sec:SM_passive_existence}

We retain the positive number-Kerr term ($\chi>0$) in Eq.~\eqref{eq:Hmain} and include cavity one-photon loss, independent spin relaxation, and dephasing, without incoherent pumping, engineered nonlinear reservoirs, or feedback. We derive the self-consistent mean-field fixed points and subsequently test their local stability to determine whether stable bright attractors arise, rather than the full finite-\(N\) Liouvillian spectrum.

The corresponding master equation is
\begin{align}
  \dot\rho={}&-i[H,\rho]+\kappa\mathcal D[a]\rho \nonumber\\
  &+\sum_{j=1}^N\left[ \Gamma_\downarrow\mathcal D[\sigma_j^-]\rho +\frac{\Gamma_\phi}{2} \mathcal D[\sigma_j^z]\rho\right].
  \label{eq:SM_passive_liouvillian}
\end{align}
Here \(\mathcal D[L]\rho=L\rho L^\dagger-\{L^\dagger L,\rho\}/2\). Photon loss removes cavity excitations, relaxation drives the emitters toward their downward states, and dephasing damps transverse coherence. The Bloch (longitudinal and transverse) decay rates are \(\Gamma_\parallel=\Gamma_\downarrow\) and \(\Gamma_\perp=\Gamma_\downarrow/2+\Gamma_\phi\). We take \(\kappa,\Gamma_\downarrow>0\) and \(\Gamma_\phi\geq0\) unless mentioned otherwise. The local spin channels need not preserve the maximal-spin sector. No such sector restriction is imposed here. 
These local jump operators define a phenomenological loss model and are not assumed to satisfy detailed balance with respect to the interacting Hamiltonian.

For any observable \(O\), Eq.~\eqref{eq:SM_passive_liouvillian} gives
\begin{equation}
\frac{d}{dt}\langle O\rangle = i\langle[H,O]\rangle +\sum_L\gamma_L \langle\mathcal D^\dagger[L]O\rangle, 
\end{equation}
where \(\mathcal D^\dagger[L]O =L^\dagger O L-\{L^\dagger L,O\}/2\), and \(\gamma_L\) denotes the corresponding dissipation rate. Applying this to the cavity operator, and using \([\hat n^2,a]=-a(2\hat n-1)\), yields
\begin{align}
  \frac{d}{dt}\langle a\rangle &=-\left(\frac{\kappa}{2}+i\omega_c\right)  \langle a\rangle - \frac{i\chi}{2N} \langle a(2\hat n-1)\rangle \nonumber\\
  &\qquad\qquad\qquad\qquad\,\,\,\,\, - \frac{2i\lambda}{N} \langle a^\dagger J_x\rangle,
  \label{eq:SM_passive_exact_a}\\
  \frac{d}{dt}\langle J_x\rangle  &=-\Gamma_\perp\langle J_x\rangle -\omega_0\langle J_y\rangle, 
  \label{eq:SM_passive_exact_Jx}\\
  \frac{d}{dt}\langle J_y\rangle &=\omega_0\langle J_x\rangle -\Gamma_\perp\langle J_y\rangle -\frac{\lambda}{N} \langle(a^2+a^{\dagger2})J_z\rangle, \label{eq:SM_passive_exact_Jy}\\
  \frac{d}{dt}\langle J_z\rangle &=\frac{\lambda}{N} \langle(a^2+a^{\dagger2})J_y\rangle -\Gamma_\parallel \left(\langle J_z\rangle+\frac N2\right).
\label{eq:SM_passive_exact_Jz}
\end{align}
These equations are exact at finite \(N\), but are not closed, as they involve higher-order cavity moments and cavity--spin correlations.

To close these equations at leading order in \(N\), define \(\alpha=\langle a\rangle/\sqrt N=X+iY\), \(s_\mu=\langle J_\mu\rangle/N\), and \(s_0=1/2\). The mean-field photon density, pair quadrature, and Kerr-shifted frequency are \(n=|\alpha|^2\), \(C=\operatorname{Re}\alpha^2=X^2-Y^2\), and \(\Omega(n)=\omega_c+\chi n\). For a symmetry-selected mean-field configuration, the leading factorization gives 
\(\langle a\hat n\rangle \simeq N^{3/2}\alpha n\),
\(\langle a^\dagger J_x\rangle \simeq N^{3/2}\alpha^*s_x\), and
\(\langle(a^2+a^{\dagger2})J_\mu\rangle \simeq2N^2Cs_\mu\), with \(\mu=y,z\).
Dividing the cavity and spin equations by \(\sqrt N\) and \(N\), respectively, and retaining the leading terms gives
\begin{align}
  \dot\alpha &=-\left[\frac{\kappa}{2}+i\Omega(n)\right]\alpha -2i\lambda s_x\alpha^*,
  \label{eq:SM_passive_cavity_flow}\\
  \dot s_x &=-\Gamma_\perp s_x-\omega_0s_y, 
  \label{eq:SM_passive_sx_flow}\\
  \dot s_y &=\omega_0s_x-\Gamma_\perp s_y-2\lambda Cs_z,
  \label{eq:SM_passive_sy_flow}\\
  \dot s_z &=2\lambda Cs_y-\Gamma_\parallel(s_z+s_0).
  \label{eq:SM_passive_sz_flow}
\end{align}
The cavity pair quadrature \(C\) drives the spin, whose polarization \(s_x\) feeds back into the cavity pair interaction. The positive Kerr term produces the density-dependent frequency shift \(\chi n\).

The normal configuration \(\alpha=s_x=s_y=0\), \(s_z=-s_0\) is always a mean-field fixed point. To find nonzero-field solutions, we first treat \(C\) as a parameter and set the three spin derivatives in Eqs.~\eqref{eq:SM_passive_sx_flow}--\eqref{eq:SM_passive_sz_flow}
\begin{align}
  s_z(C)&=-\frac{s_0}{1+\eta_{\rm s}C^2},
  \label{eq:SM_passive_sz_solution}\\
  s_x(C)&=\frac{2\lambda\omega_0 C}{\omega_0^2+\Gamma_\perp^2}\,s_z(C),
  \label{eq:SM_passive_sx_solution} \\
  s_y(C)&=-\frac{\Gamma_\perp}{\omega_0}s_x(C),
  \label{eq:SM_passive_sy_solution}
\end{align}
where \(\eta_{\rm s}=4\lambda^2\Gamma_\perp/[\Gamma_\parallel(\omega_0^2+\Gamma_\perp^2)]\). The denominator \(1+\eta_{\rm s}C^2\) describes spin saturation. 

The value of \(C\) must also satisfy the stationary cavity equation. For \(\alpha=\sqrt n\,e^{i\phi}\neq0\), Eq.~\eqref{eq:SM_passive_cavity_flow} gives \((\kappa/2+i\Omega)e^{2i\phi}=-2i\lambda s_x\). Its modulus and real part, together with \(C=n\cos(2\phi)\), imply
\begin{align}
  D_c(n)&=4\lambda^2s_x^2(C), \\
  2\lambda s_x(C)C&=-n\Omega(n),
\label{eq:SM_passive_phase_condition}
\end{align}
where \(D_c(n)=\Omega(n)^2+\kappa^2/4\). The first condition balances pair coupling against detuning and loss, while the second constrains the cavity phase. Together they give \(C^2=n^2\Omega(n)^2/D_c(n)\). Substituting Eqs.~\eqref{eq:SM_passive_sz_solution} and~\eqref{eq:SM_passive_sx_solution}, and eliminating \(C\), yields (\(n, \Omega(n)>0\))
\begin{equation}
\begin{aligned}
0={}&(\omega_0^2+\Gamma_\perp^2)D_c(n)+\frac{4\lambda^2\Gamma_\perp}
{\Gamma_\parallel}n^2\Omega(n)^2\\
&\qquad\qquad-4\lambda^2\omega_0s_0\,n\Omega(n).
\end{aligned}
\label{eq:SM_passive_scalar_fixed_point}
\end{equation}
For \(\Omega(n)=\omega_c+\chi n\), this is a quartic equation for \(n\). For each positive real root, set \(C=\pm n\Omega(n)/\sqrt{D_c(n)}\) and obtain the spin components from Eqs.~\eqref{eq:SM_passive_sz_solution}--\eqref{eq:SM_passive_sy_solution}. The cavity phase follows from
\begin{equation}
  e^{2i\phi} =\frac{-2i\lambda s_x(C)} {\kappa/2+i\Omega(n)}.
  \label{eq:SM_passive_phase_solution}
\end{equation}
The two signs of \(C\), each allowing two phases differing by \(\pi\), yield four \(\mathbb Z_4\)-related fixed points for each admissible positive density.

Since \(J_x=\frac12\sum_{j=1}^N\sigma_j^x\), \(|s_x|=|\langle J_x\rangle|/N\leq1/2\). The stationary cavity condition requires \(D_c(n)\leq\lambda^2\). For \(\chi>0\), a nonzero fixed point therefore requires \(\lambda^2>\omega_c^2+\kappa^2/4\) and obeys
\begin{equation}
  n\leq \frac{\sqrt{\lambda^2-\kappa^2/4}-\omega_c}{\chi}.
  \label{eq:SM_passive_amplitude_cap}
\end{equation}
This necessary bound limits the stationary photon density but does not guarantee a physical solution.

\subsection{Local stability of the passive fixed points}
\label{sec:SM_passive_stability}

A fixed point is locally attracting only if small perturbations decay. For \(\mathbf u=(X,Y,s_x,s_y,s_z)^T\), the mean-field equations read \(\dot{\mathbf u}=\mathbf F(\mathbf u)\), with
\begin{equation}
\mathbf F=
\begin{pmatrix}
-\frac{\kappa}{2}X+[\Omega(n)-2\lambda s_x]Y\\
-\frac{\kappa}{2}Y-[\Omega(n)+2\lambda s_x]X\\
-\Gamma_\perp s_x-\omega_0s_y\\
\omega_0s_x-\Gamma_\perp s_y-2\lambda Cs_z\\
2\lambda Cs_y-\Gamma_\parallel(s_z+s_0)
\end{pmatrix}.
\label{eq:SM_passive_real_flow}
\end{equation}
Here \(n\) and \(C\) evolve with the cavity field.

For a small perturbation \(\mathbf u=\mathbf u_*+\delta\mathbf u\) around a fixed point \(\mathbf u_*\), the linearized dynamics is \(\delta\dot{\mathbf u} =\mathcal J_*\delta\mathbf u\), where \(\mathcal J_*= [\partial\mathbf F/\partial\mathbf u]_{\mathbf u_*}\). Explicitly,
\begin{equation}
{\tiny
\mathcal J_*\!=\!
\left.
\begin{pmatrix}
-\frac{\kappa}{2}+2\chi XY
&\Omega-2\lambda s_x+2\chi Y^2
&-2\lambda Y&0&0\\
-\Omega-2\lambda s_x-2\chi X^2
&-\frac{\kappa}{2}-2\chi XY
&-2\lambda X&0&0\\
0&0&-\Gamma_\perp&-\omega_0&0\\
-4\lambda Xs_z&4\lambda Ys_z
&\omega_0&-\Gamma_\perp&-2\lambda C\\
4\lambda Xs_y&-4\lambda Ys_y
&0&2\lambda C&-\Gamma_\parallel
\end{pmatrix}
\right|_{\mathbf u_*}.
}
\label{eq:SM_passive_jacobian}
\end{equation}
Let us define
\begin{equation}
  \mu_{\max} =\max_j\operatorname{Re}\mu_j(\mathcal J_*).
  \label{eq:SM_mumax_def}
\end{equation}
The fixed point is locally attracting for \(\mu_{\max}<0\) and unstable for \(\mu_{\max}>0\). At \(\mu_{\max}=0\), this linear sign test fails and analysis beyond linear order is required.

The normal fixed point can be classified analytically. The pair interaction enters only beyond linear order about this configuration, so the Jacobian is independent of \(\lambda\), with eigenvalues
\begin{equation}
-\frac{\kappa}{2}\pm i\omega_c,\qquad
-\Gamma_\perp\pm i\omega_0,\qquad
-\Gamma_\parallel.
\label{eq:SM_normal_eigenvalues}
\end{equation}
All real parts are negative for positive damping rates. Thus, the normal fixed point is locally attracting for any \(\lambda\), although additional attractors may coexist.

At nonzero fixed points, cavity and spin perturbations remain coupled.  In particular, \(\delta n=2X_*\delta X+2Y_*\delta Y\) and \(\delta C=2X_*\delta X-2Y_*\delta Y\); neither quantity is held fixed. The spin must be perturbed independently, rather than assumed to follow its stationary response instantaneously. The Jacobian has a fifth-degree characteristic polynomial whose coefficients depend on the stationary configuration and decay rates. We reconstruct each positive-density solution of Eq.~\eqref{eq:SM_passive_scalar_fixed_point} and test its full spectrum. Symmetry-related configurations have identical Jacobian spectra, so one representative suffices for each density. The following specialization permits additional analytical conclusions.

\subsection{Existence and stability of bright states}
\label{sec:SM_passive_unequal_rates}

To obtain explicit analytical results and compare directly with the closed-system phase diagram, we set \(\omega_c=\omega_0=\chi=1\). While \(\omega_c=1\) fixes the frequency unit, the remaining equalities select a representative Hamiltonian parameter set; the resulting quantitative thresholds may differ for other frequency and Kerr ratios.

We keep \(\kappa>0\) and the microscopic spin rates \(\Gamma_\downarrow>0\), \(\Gamma_\phi\geq0\) independently variable. We first classify finite-density fixed points and then treat the zero-spin-decay limits separately.

\paragraph{Stationary-coupling relation and density bound.}
To determine the allowed stationary photon densities, define \(r_{\rm s}=\Gamma_\perp/\Gamma_\parallel = 1/2+\Gamma_\phi/\Gamma_\downarrow\geq1/2\), for \(\Gamma_\downarrow>0\). Rather than solving the quartic Eq.~\eqref{eq:SM_passive_scalar_fixed_point} explicitly, we express the coupling required to sustain a given density \(n>0\) as
\begin{equation}
  \lambda^2=\Lambda(n) \equiv(1+\Gamma_\perp^2)F_{\kappa,r_{\rm s}}(n),
\label{eq:SM_passive_unequal_stationary}
\end{equation}
where
\begin{equation}
  F_{\kappa,r_{\rm s}}(n) = \frac{(1+n)^2+\kappa^2/4} {2n(1+n)[1-2r_{\rm s}n(1+n)]}.
  \label{eq:SM_F_k_rs_n}
\end{equation}
For fixed decay rates, the intersections of \(\Lambda(n)\) with a prescribed \(\lambda^2\) give the possible stationary photon densities. Since the numerator is positive, a physical solution requires \(1-2r_{\rm s}n(1+n)>0\), yielding
\begin{equation}
  0<n<n_{\rm cap}(r_{\rm s}) =\frac{\sqrt{1+2/r_{\rm s}}-1}{2}.
  \label{eq:SM_passive_unequal_cap}
\end{equation}
Furthermore, \(2y(1-2r_{\rm s}y)\leq1/(4r_{\rm s})\), with \(y=n(1+n)\), and \((1+n)^2>1\) give the necessary condition
\begin{equation}
  \lambda^2 > 4r_{\rm s}(1+\Gamma_\perp^2) \left(1+\frac{\kappa^2}{4}\right).
  \label{eq:SM_passive_rate_restriction}
\end{equation}
These inequalities restrict the allowed density and decay rates at a given coupling, but do not guarantee the existence of a nonzero fixed point.

\begin{figure*}[t]
\centering
\includegraphics[width=\textwidth,height=4.5in]
{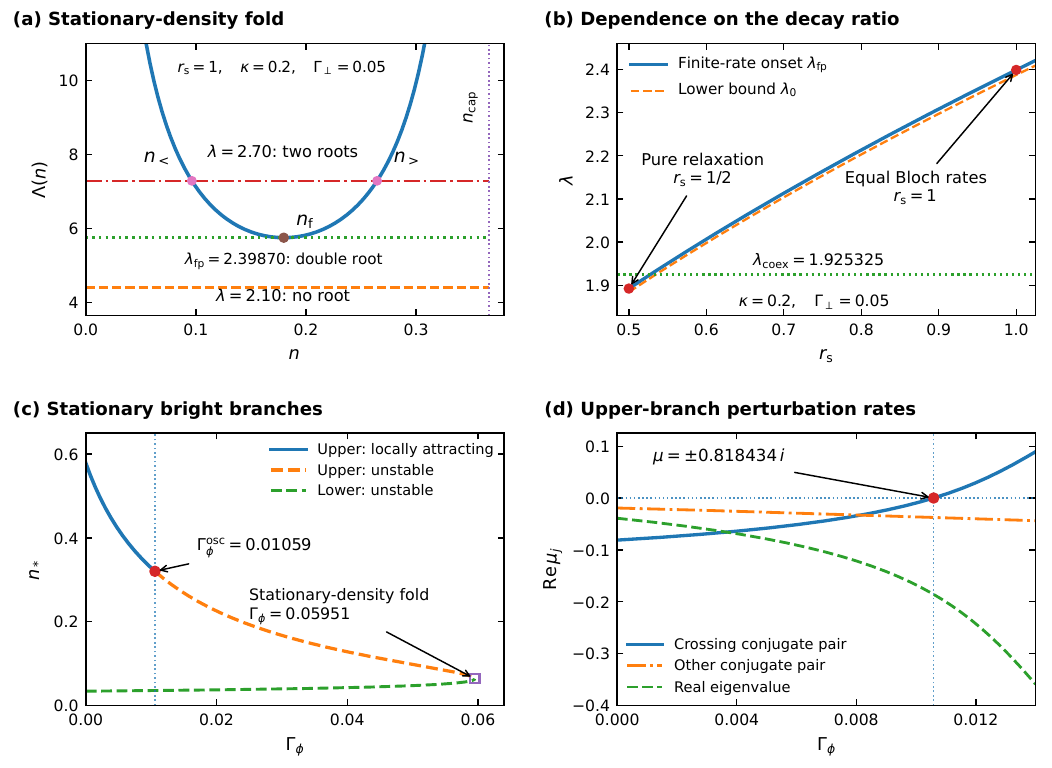}\\
\vspace{-3ex}
\caption{\textbf{Existence and local stability of bright fixed points.} All panels use \(\omega_c=\omega_0=\chi=1\) and \(\kappa=0.2\).
(a) Stationary-coupling curve at \(r_{\rm s}=1\), \(\Gamma_\perp=0.05\); horizontal lines illustrate zero, double, and two positive roots, and the vertical dotted line marks \(n_{\rm cap}\).
(b) Finite-rate threshold \(\lambda_{\rm fp}\), lower bound \(\lambda_0\), and Hamiltonian coexistence, with \(\Gamma_\perp=0.05\) and \(\Gamma_\parallel=\Gamma_\perp/r_{\rm s}\).
(c) Bright stationary densities at \(\lambda=4\), \(\Gamma_\downarrow=0.02\), with solid/dashed segments denoting locally attracting/unstable fixed points. The circle marks loss of attraction; the square marks the stationary-density fold.
(d) Real parts of all five upper-branch Jacobian eigenvalues near the oscillatory crossing; each conjugate pair has coincident real parts. Each stationary density represents a symmetry-related quartet.}
\label{fig:SM_passive_fold_thresholds}
\end{figure*}

\paragraph{Existence of nonzero fixed points.}
We now determine the coupling at which stationary
bright configurations first appear. The function \(F_{\kappa,r_{\rm s}}\) diverges at both ends of \(0<n<n_{\rm cap}\). For \(r_{\rm s}\geq1/2\), the numerator of its derivative has a negative constant term and nonnegative positive-power coefficients, with strictly positive coefficients from \(n^2\) onward. It increases through zero once in the physical interval, so \(F_{\kappa,r_{\rm s}}\) has a unique minimum at \(n=n_{\rm f}\). This defines the existence threshold
\begin{equation}
  \lambda_{\rm fp}^2 =(1+\Gamma_\perp^2)F_{\kappa,r_{\rm s}}(n_{\rm f}).
  \label{eq:SM_passive_unequal_fold}
\end{equation}
Below \(\lambda_{\rm fp}\) there is no positive stationary density. At \(\lambda_{\rm fp}\), there is one double root, and above it two densities \(n_< < n_>\) [see Fig.~\ref{fig:SM_passive_fold_thresholds}(a)]. Each density specifies a \(\mathbb Z_4\)-related quartet, not necessarily an attractor.

At fixed \(r_{\rm s}\), the minimizer \(n_{\rm f}\) depends on \(\kappa\), but not on the overall spin-decay scale \(1+\Gamma_\perp^2\). Since \(F_{\kappa,r_{\rm s}}\geq F_{0,r_{\rm s}}\) and
\(1+\Gamma_\perp^2\geq1\),
\begin{align}
  \lambda_{\rm fp}^2 &=(1+\Gamma_\perp^2)\min_n F_{\kappa,r_{\rm s}}(n) \nonumber\\
  &\geq\min_n F_{0,r_{\rm s}}(n) \equiv\lambda_0^2(r_{\rm s}).
  \label{eq:SM_passive_unequal_threshold_bound}
\end{align}
Both minima are over \(0<n<n_{\rm cap}\). For \(F_{0,r_{\rm s}}(n)=(1+n)/[2n(1-2r_{\rm s}n(1+n))]\), the condition \(F_{0,r_{\rm s}}'(n_0)=0\) gives
\begin{equation}
  4r_{\rm s}n_0(1+n_0)^2=1,\qquad
  \lambda_0^2(r_{\rm s})=\frac{(1+n_0)^2}{n_0(1+2n_0)}.
\end{equation}
The bound is approached as
\(\kappa,\Gamma_\parallel,\Gamma_\perp\to0^+\) at fixed \(r_{\rm s}\). It is not the stationary problem with zero spin damping.

Pure relaxation (\(r_{\rm s}=1/2\)) gives \(\lambda_0\simeq1.884573\), whereas equal Bloch rates (\(r_{\rm s}=1\)) give \(\lambda_0\simeq2.387149\). Thus, equal rates exclude nonzero fixed points at the Hamiltonian coexistence coupling \(\lambda_{\rm coex}\simeq1.925325\), but the bound alone does not exclude them for pure relaxation. Figure~\ref{fig:SM_passive_fold_thresholds}(b) compares the limiting and finite-rate thresholds and gives a pure-relaxation example with \(\lambda_{\rm fp}<\lambda_{\rm coex}\). Table~\ref{tab:SM_passive_physical_rates} converts the model rates of panels (a,b) to illustrative physical units.

\paragraph{Local stability.}
The stationary densities found above specify possible bright configurations; we now determine whether perturbations around them decay. For either nonzero branch, define \(p_*(z)=\det(z\mathbf1-\mathcal J_*)\) using the full cavity--spin Jacobian. Its zeros are the perturbation growth rates, so any zero with positive real part signals instability. Evaluating its constant term gives
\begin{equation}
\begin{aligned}
  p_*(0)={}&8\Gamma_\parallel n^2(1+n)\\
  &\times[1-2r_{\rm s}n(1+n)]\Lambda'(n),
\end{aligned}
\label{eq:SM_passive_unequal_slope_stability}
\end{equation}
where \(\Lambda'(n)\) is taken at fixed decay rates. The prefactor is positive in the allowed density interval. Thus the descending branch, \(n_<\), has \(p_*(0)<0\). Since \(p_*(z)\to+\infty\) as real \(z\to+\infty\), it must have a positive real zero, corresponding to a growing perturbation. The lower-density quartet is therefore unstable throughout this positive-relaxation parameter family. On the upper branch, \(p_*(0)>0\) is necessary but not sufficient for attraction; all five eigenvalues must have negative real parts.

To illustrate this distinction, we vary pure dephasing at \(\lambda=4\), \(\kappa=0.2\), and \(\Gamma_\downarrow=0.02\), so that \(\Gamma_\perp=0.01+\Gamma_\phi\). Figure~\ref{fig:SM_passive_fold_thresholds}(c) follows both stationary densities and classifies them using the full Jacobian. The upper branch is locally attracting at weak dephasing but loses stability at \(\Gamma_\phi^{\rm osc}\simeq0.0105878\). At this boundary, a conjugate pair crosses the imaginary axis at \(\mu=\pm0.818434i\), while the other three eigenvalues retain negative real parts [Fig.~\ref{fig:SM_passive_fold_thresholds}(d)]. 
The upper branch remains as an unstable stationary solution until it merges with the lower branch at \(\Gamma_\phi^{\rm fold}\simeq0.0595145\). Dephasing therefore destroys attraction before eliminating the stationary bright configurations. The nonzero crossing frequency identifies a linear oscillatory instability, not the existence or stability of a periodic orbit.

\paragraph{Zero-rate limits.}
With \(\kappa>0\), pure dephasing (\(\Gamma_\parallel=0\), \(\Gamma_\perp=\Gamma_\phi>0\)) must be treated separately since \(r_{\rm s}\) is undefined. A nonzero cavity fixed point requires \(s_xC\neq0\). However, \(\dot s_z=2\lambda Cs_y=0\) implies \(s_y=0\), and \(\dot s_x=-\Gamma_\perp s_x-s_y=0\) then forces \(s_x=0\) (a contradiction). Thus, no finite-field fixed point exists. Instead, \(\alpha=s_x=s_y=0\) with arbitrary \(s_z\in[-1/2,1/2]\) forms a continuous family with a neutral longitudinal direction.

When both spin-dissipation channels vanish, \(\Gamma_\downarrow=\Gamma_\phi=0\), the mean-field spin length is conserved, so stationary solutions depend on its initial value. This problem is not obtained by substituting zero rates into Eq.~\eqref{eq:SM_passive_unequal_stationary}. The alternative \(\Gamma_\perp=0<\Gamma_\parallel\) is incompatible with the specified Lindblad channels.

Hamiltonian confinement, existence of stationary bright fields, and local attraction are therefore distinct. The latter two properties depend on cavity loss and spin dissipation, and the representative scan shows that attraction can be lost before the fixed points disappear. These conclusions concern local mean-field dynamics, not global basins, subsequent nonlinear attractors, or the finite-\(N\) Liouvillian spectrum.

\begin{table}[t]
\caption{\textbf{Illustrative rate conversion for panels (a,b) of Fig.~\ref{fig:SM_passive_fold_thresholds}.} At \(\omega_{\rm ref}/2\pi=3\,\mathrm{MHz}\), the physical rate is \(\gamma=d\omega_{\rm ref}\) and \(\tau=1/\gamma\). These are assumed scales, not measured device parameters. For cavity loss, \(\tau\) is the photon-number lifetime; the field-amplitude lifetime is \(2\tau\).}
\label{tab:SM_passive_physical_rates}
\centering
\setlength{\tabcolsep}{4pt}
\renewcommand{\arraystretch}{1.08}
\begin{ruledtabular}
\begin{tabular}{@{}lccc@{}}
Rate & \(d\) & \(\gamma/2\pi\) (kHz)
& \(\tau\) (\(\mu\)s)\\
\hline
\multicolumn{4}{l}{Shared cavity loss and transverse spin decay}\\
\(\kappa\) & 0.200 & 600 & 0.265\\
\(\Gamma_\perp\) & 0.050 & 150 & 1.061\\[2pt]
\multicolumn{4}{l}{Pure relaxation: \(r_{\rm s}=1/2\)}\\
\(\Gamma_\downarrow\) & 0.100 & 300 & 0.531\\
\(\Gamma_\phi\) & 0 & 0 & \(\infty\)\\[2pt]
\multicolumn{4}{l}{Equal Bloch rates: \(r_{\rm s}=1\)}\\
\(\Gamma_\downarrow\) & 0.050 & 150 & 1.061\\
\(\Gamma_\phi\) & 0.025 & 75 & 2.122\\
\end{tabular}
\end{ruledtabular}
\end{table}


\section{Circuit-QED emulation and \texorpdfstring{finite-$N$}{finite-N} benchmarks}
\label{sec:SM_circuit_QED}

We construct a driven rotating-frame emulator of
Eq.~\eqref{eq:Hmain} and distinguish the target Hamiltonian
from finite-drive corrections and illustrative noise models.
The construction concerns finite-$N$ spectroscopy and coherent
branch transfer, not an undriven realization or a passive bright
attractor. System coefficients and decay
rates are quoted as angular frequencies unless divided by $2\pi$.

\subsection{Circuit ingredients and native pair coupling}
\label{subsec:SM_circuit_native}

Consider a nonlinear storage mode $a$ coupled to $N$ qubits,
with the effective low-energy Hamiltonian ($\hbar=1$)
\begin{align}
H_{\rm lab}={}&\omega_r\hat n
 +\frac{\mathcal K_N}{2}a^{\dagger2}a^2
 +\sum_{k=1}^{N}\frac{\omega_{qk}}{2}\sigma_z^{(k)}
\nonumber\\
&+\sum_{k=1}^{N}g_{2k}\sigma_x^{(k)}(a+a^\dagger)^2
 +H_{\rm d}(t)+H_{\rm par}.
\label{eq:SM_circuit_lab}
\end{align}
Here $\mathcal K_N>0$ is the \emph{net} storage self-Kerr
coefficient, with the positive sign shown explicitly.
Quarton coupling can supply a positive contribution opposing
transmon anharmonicity, and near-cancellation of self-Kerr has
been demonstrated~\cite{Ye2025Quarton}. Realizing the required net
positive Kerr together with all pair couplings remains a
full-network design requirement.

An asymmetric SQUID provides a route to the pair
interaction~\cite{Ayyash2024}. Its odd Josephson potential contains
\begin{align}
-E_s\sin(\varphi_q-\varphi_r)
={}&\cdots+\frac{E_s}{2}\varphi_q\varphi_r^2
 -\frac{E_s}{2}\varphi_q^2\varphi_r
\nonumber\\
&-\frac{E_s}{6}\varphi_r^3+O(\varphi^5).
\label{eq:SM_odd_expansion}
\end{align}
Writing $\varphi_q=\eta_qX_q$ and
$\varphi_r=\eta_r(a+a^\dagger)$, with the qubit transition
matrix element included in $\eta_q$ so that
$X_q\mapsto\sigma_x$, gives
\begin{equation}
g_2=\frac{E_s\eta_q\eta_r^2}{2}.
\label{eq:SM_microscopic_g2}
\end{equation}
The same coupler also generates unwanted terms. For one qubit,
a representative remainder after projection and displacement is
\begin{align}
H_{\rm par}={}&-\widetilde g_4(a+a^\dagger)^3
 -\widetilde g_5\sigma_z(a+a^\dagger)
\nonumber\\
&-\widetilde g_c(\sigma_+-\sigma_-)(a^\dagger-a)
 +H_{\rm off}+O(\varphi^5),
\label{eq:SM_circuit_parasitics}
\end{align}
where $\sigma_\pm=(\sigma_x\pm i\sigma_y)/2$ and
$H_{\rm off}$ collects additional offsets and off-resonant terms.
Their coefficients require full-network quantization; the pair
term is not generated in isolation. In particular, higher powers
of $X_q$ must be projected as complete operators, rather
than inferred by squaring the projected matrix $\sigma_x$.

For uniform couplings $g_{2k}=g_N$ and nearly identical qubits
with $\omega_q\simeq2\omega_r$, a rotating-wave approximation
(RWA) retains two-photon exchange:
\begin{align}
H_{\rm 2JC}={}&\omega_r\hat n
 +\frac{\mathcal K_N}{2}\hat n(\hat n-1)+\omega_qJ_z
\nonumber\\
&+g_N(J_+a^2+J_-a^{\dagger2}),
\label{eq:SM_nativeJC}
\end{align}
where $J_\pm=J_x\pm iJ_y=\sum_k\sigma_\pm^{(k)}$.
Discarding the other terms requires their matrix elements to be
small compared with their oscillation frequencies over the
occupied photon range. Residual dispersive shifts must be
calibrated; their occupation-dependent parts cannot generally
be absorbed into a single detuning. All numerical benchmarks
below start after this first RWA and do not test
$H_{\rm par}$ or the full Josephson expansion.

\subsection{Two-tone dressing and target-Hamiltonian map}
\label{subsec:SM_circuit_floquet_map}

Drive the qubits uniformly with phase-coherent tones of amplitudes
$\Omega_1,\Omega_2$ and frequencies $\omega_{d1},\omega_{d2}$.
After the carrier RWA, a frame rotating the qubits at
$\omega_{d1}$ and the cavity at $\omega_{d1}/2$ gives
\begin{align}
H_1(t)={}&\delta_2\hat n
 +\frac{\mathcal K_N}{2}\hat n(\hat n-1)
 +\Delta J_z+\Omega_1J_x
\nonumber\\
&+g_N(J_+a^2+J_-a^{\dagger2})
\nonumber\\
&+\frac{\Omega_2}{2}
 \left[e^{i(\delta_dt+\phi_2)}J_++\mathrm{H.c.}\right],
\label{eq:SM_first_drive_frame}
\end{align}
where \(\Delta=\omega_q-\omega_{d1}\), 
\(\delta_2=\omega_r-\frac{\omega_{d1}}{2}\),
\(\delta_d=\omega_{d1}-\omega_{d2}\).
Choose $\Delta=0$, $\delta_d=\Omega_1$, and $\phi_2=0$.
In the interaction picture of $\Omega_1J_x$, period averaging
then yields~\cite{Ayyash2024}
\begin{equation}
H_{\rm av}=\delta_2\hat n
 +\frac{\mathcal K_N}{2}\hat n(\hat n-1)
 +\frac{\Omega_2}{2}J_z+g_N(a^2+a^{\dagger2})J_x.
\label{eq:SM_circuit_average_H}
\end{equation}
The strong drive supplies balanced co- and counter-rotating pair
terms, while the weaker tone supplies the effective spin splitting.
The basic hierarchy is
\begin{equation}
\Omega_1\gg\Omega_2,|\Delta|,\qquad
\Omega_\ell\ll\omega_{d\ell}\quad(\ell=1,2).
\label{eq:SM_strong_drive_conditions}
\end{equation}
Averaging also requires off-resonant, perturbative oscillatory
couplings on the populated subspace. These involve both the
photon factor $\sqrt{(n_a+1)(n_a+2)}$ and collective-spin
matrix elements; the Kerr-shifted pair transition is
$2\delta_2+\mathcal K_N(2n_a+1)$.
A hierarchy stated only in terms of bare $g_N$ is therefore
not a uniform bound in photon number or $N$.

Within the averaged Hamiltonian, the parameter map is exact:
\begin{equation}
\begin{aligned}
\mathcal K_N=\frac{\chi}{N},\,\,
\delta_2=\omega_c+\frac{\chi}{2N},\,\,
\Omega_2=2\omega_0,\,\,
g_N=\frac{\lambda}{N}.
\end{aligned}
\label{eq:SM_circuit_parameter_map}
\end{equation}
The detuning compensates the normally ordered Kerr term,
\begin{equation}
\delta_2\hat n+\frac{\mathcal K_N}{2}\hat n(\hat n-1)
=\omega_c\hat n+\frac{\chi}{2N}\hat n^2.
\label{eq:SM_normal_ordering_compensation}
\end{equation}
The coupling convention is
\begin{equation}
g_N(a^2+a^{\dagger2})J_x
=\frac{\lambda}{2N}\sum_{k=1}^{N}
(a^2+a^{\dagger2})\sigma_x^{(k)}.
\label{eq:SM_per_emitter_factor}
\end{equation}
Thus the native exchange coefficient is $g_N=\lambda/N$,
whereas each dressed single-qubit term has coefficient
$\lambda/(2N)$. The intended size sequence requires retuning
both $\mathcal K_N$ and $g_N$; it is not obtained by increasing
$N$ at fixed circuit couplings.

\subsection{\texorpdfstring{Finite-$N$}{Finite-N} targets and averaging errors}
\label{subsec:SM_circuit_targets}

Table~\ref{tab:SM_circuit_targets} gives target-Hamiltonian
benchmarks for $(\omega_c,\omega_0,\chi)/2\pi=(3,3,3)$ MHz.
The branch-crossing coupling, same-sector gap, and branch states
$|{\rm N}\rangle,|{\rm B}\rangle$ are defined as in
Sec.~\ref{sec:SM_symmetry_gap}.
Writing $\Delta\bar n=N\Delta n_N$, ideal evolution from
$|{\rm N}\rangle$ at the branch crossing gives
$P_{\rm B}(t)=\sin^2[\Delta_{\rm NB}^{(0)}t/2]$ and
$t_{\rm swap}=\pi/\Delta_{\rm NB}^{(0)}$.
The table specifies the target energy scales and occupations
that an implementation must reproduce.

\begin{table}[t]
\caption{Finite-$N$ target parameters and ideal branch-transfer
scales for $(\omega_c,\omega_0,\chi)/2\pi=(3,3,3)$ MHz.
All frequency columns are ordinary MHz. The final single-qubit
pair coefficient is $g_N/2$; $\Delta\bar n$ is the photon-number
contrast. All values are rounded numerical results for the target
Hamiltonian.}
\label{tab:SM_circuit_targets}
\centering
\begin{ruledtabular}
\begin{tabular}{ccccccc}
$N$ & $\lambda_N^\times/2\pi$ & $\mathcal K_N/2\pi$ &
$g_N/2\pi$ & $\Delta_{\rm NB}^{(0)}/2\pi$ &
$\Delta\bar n$ & $t_{\rm swap}$ ($\mu$s)\\
\hline
4 & 6.2862 & 0.750 & 1.5716 & 2.2295 & 3.618 & 0.224\\
6 & 6.0307 & 0.500 & 1.0051 & 1.2852 & 4.786 & 0.389\\
8 & 5.9347 & 0.375 & 0.7418 & 0.7764 & 6.120 & 0.644
\end{tabular}
\end{ruledtabular}
\end{table}

Omitting the detuning compensation shifts the target crossings by
\begin{equation}
\frac{\delta\lambda_N^\times}{2\pi}
=(-377.9,-255.0,-192.6)~\mathrm{kHz},
\quad N=(4,6,8).
\label{eq:SM_uncompensated_crossing_error}
\end{equation}
To distinguish this static correction from averaging errors, set
$\vartheta=\Omega_1t$ and $\mathcal A=a^2-a^{\dagger2}$.
Under the drive choices above, the exact toggling-frame
Hamiltonian within the first-RWA model is
\begin{align}
H_{\rm tog}(t)=H_{\rm av}
 &+(\Omega_2J_x+ig_N\mathcal A J_y)\cos\vartheta -ig_N\mathcal A J_z\sin\vartheta
\nonumber\\
&\,\, -\frac{\Omega_2}{2}
 [J_z\cos2\vartheta+J_y\sin2\vartheta].
\label{eq:SM_exact_toggling_H}
\end{align}
For the Fourier convention $H_{\rm tog}(t)=\sum_m V_m
 e^{im\Omega_1t}$, the first van Vleck correction is
$\sum_{m>0}[V_m,V_{-m}]/(m\Omega_1)$, yielding
\cite{EckardtAnisimovas2015}
\begin{align}
H_{\rm vV}^{(1)}={}&\frac{\Omega_2^2}{16\Omega_1}J_x
 -\frac{i\Omega_2g_N}{2\Omega_1}\mathcal A J_y
-\frac{g_N^2}{2\Omega_1}\mathcal A^2J_x.
\label{eq:SM_van_Vleck_correction}
\end{align}
The 
$J_x$ term can be compensated by dressed-field
calibration; the remaining terms are occupation-dependent
and need not preserve the target \(\mathbb Z_4\) symmetry.
On a selected real or imaginary bright configuration,
$\langle\mathcal A\rangle=0$, but its fluctuations contribute
to finite-size energies. This branch-local observation does not
establish unchanged thermodynamics of the full driven system
at fixed $\Omega_1$.

For orientation, evaluating the compensated correction in a
real product-CS bright configuration at thermodynamic coexistence
gives the leading estimate
\begin{equation}
\delta\lambda_{N,\rm CS}^{\times}
\simeq-\frac{\lambda_{\rm coex}^2}{N\Omega_1}.
\label{eq:SM_Floquet_crossing_shifts}
\end{equation}
At $\Omega_1/\omega_c=100$, this gives
$(-27.8,-18.5,-13.9)$ kHz after division by $2\pi$, for
$N=(4,6,8)$. These are product-CS estimates, not exact
finite-$N$ Floquet crossings; correlated fluctuations and
higher-order drive corrections are not included.

We also compare propagation under $H_{\rm tog}(t)$ with
$U_{\rm av}(t)=e^{-itH_{\rm av}}$ over
$T=2\pi/\Omega_1$. For the six lowest target eigenstates
$|\psi_s\rangle$ in the maximal-spin space (all four $q$
sectors included), use
\begin{equation}
F_s=|\langle\psi_s|U_{\rm av}^\dagger(T)
 U_{\rm tog}(T)|\psi_s\rangle|^2.
\label{eq:SM_one_period_fidelity}
\end{equation}
At $\Omega_1/\omega_c=100$ and photon cutoffs $24,26,28$
for $N=4,6,8$, respectively, the maximum $1-F_s$ is below
$3\times10^{-4}$.
This tests eigenstate survival over one period, not relative
phase accuracy, channel fidelity, or long-time agreement.
A complementary transfer test starts in $|{\rm N}\rangle$
and monitors $P_{\rm B}(t)=|\langle{\rm B}|\psi(t)\rangle|^2$.
For the explicitly calibrated model
$H_{\rm cal}(t)=H_{\rm tog}(t)-h_xJ_x$, with
$h_x=0.01\omega_c$, the first transfer peak satisfies
\(P_{\rm B}^{\max}>0.997\) 
using the same cutoffs.
These finite-model checks do not
include the discarded laboratory terms or multilevel leakage.

\subsection{Dressed noise and open-system scope}
\label{subsec:SM_circuit_open_pulse}

The strong tone transforms the noise channels as well as the
Hamiltonian. With
$\mathcal D[L]\rho=L\rho L^\dagger-\tfrac12\{L^\dagger L,\rho\}$,
a single-qubit laboratory-frame model is
$\mathcal L_q=\gamma_1\mathcal D[\sigma_-]
 +(\gamma_\phi/2)\mathcal D[\sigma_z]$.
Averaging its rotation about $x$ in the white-noise Markov limit
gives
\begin{equation}
\overline{\mathcal L}_q
=\frac{\gamma_1}{4}\mathcal D[\sigma_x]
 +\left(\frac{\gamma_1}{8}+\frac{\gamma_\phi}{4}\right)
 [\mathcal D[\sigma_y]+\mathcal D[\sigma_z]].
\label{eq:SM_dressed_noise}
\end{equation}
For colored noise, the dressed rates require spectral information
at the relevant drive sidebands; the bare rates alone are
insufficient. Spin-lock measurements under the applied tone
would characterize this requirement.

The illustrative $N=4$ open-system calculation evolves
\begin{equation}
\dot\rho=-i[H_{\rm av},\rho]+\kappa\mathcal D[a]\rho
 +\sum_{k=1}^{4}\mathcal L_q^{(k)}\rho,
\label{eq:SM_open_master_equation}
\end{equation}
starting from $|{\rm N}\rangle\langle{\rm N}|$ at
$\lambda_4^\times$, and measures
$P_{\rm B}(t)=\langle{\rm B}|\rho(t)|{\rm B}\rangle$.
We compare inserting the unrotated channels directly in the target
frame with using $\overline{\mathcal L}_q$.
Local noise can change total spin, so these evolutions retain
all $2^4$ spin states rather than restricting to $j=2$.
For identical channels and permutation-invariant initial data,
permutational invariance reduces the \emph{spin operator space}
from $4^4=256$ to $\binom{7}{3}=35$
components~\cite{Shammah2018}.
At $n_a^{\max}=10$, comparison with the unreduced Liouvillian
agrees within $4.0\times10^{-15}$ in branch populations and photon
number. This checks the symmetry reduction at the same cutoff,
not convergence with respect to that cutoff.

For $\kappa/2\pi=25$ kHz and
$\gamma_1/2\pi=\gamma_\phi/2\pi=5$ kHz, the target-frame
and dressed-noise benchmarks are
\begin{equation}
P_{{\rm B},{\rm target}}^{\max}=0.9242,
\qquad P_{{\rm B},{\rm dressed}}^{\max}=0.9152.
\label{eq:SM_open_pulse_base_results}
\end{equation}
These use photon cutoffs $24$ and $20$, respectively.
Adding phenomenological effective-frame dephasing
$(\gamma_p/2)\sum_k\mathcal D[\sigma_z^{(k)}]$ to the dressed
model gives
\begin{equation}
P_{\rm B}^{\max}=(0.9080,0.9010,0.8869,0.8468)
\label{eq:SM_open_pulse_pump_dephasing}
\end{equation}
for $\gamma_p/2\pi=(5,10,20,50)$ kHz.
These benchmarks quantify transient transfer under assumed
Markov rates, not passive bright-state attraction. They do not
combine full-circuit, finite-drive, and multilevel noise effects.

As one parameter illustration, taking $\omega_r/2\pi=5.000$ GHz
and $N=6$ gives
\begin{equation}
\begin{aligned}
(\mathcal K_6,\delta_2)/2\pi&=(0.500,3.250)~\mathrm{MHz},\\
(\Omega_2,\Omega_1)/2\pi&=(6.000,150\text{--}300)~\mathrm{MHz}.
\end{aligned}
\label{eq:SM_N6_operating_point}
\end{equation}
Then $\omega_{d1}/2\pi=9.9935$ GHz,
$\omega_q\simeq\omega_{d1}$, and
$\omega_{d2}=\omega_{d1}-\Omega_1$ before residual calibration.
These frequencies specify the effective map, but do not check
the qubit anharmonicity or leakage under strong driving.

A finite-$N$ test could use photon-number-sensitive readout of
branch transfer or a weak cavity-detuning modulation proportional
to $\hat n$ to address the same-sector avoided gap.
A linear cavity probe changes generalized parity and does not
directly connect the two $q=0$ levels.
Establishing quantitative experimental performance additionally
requires full-network quantization, multilevel driven dynamics,
measured dressed noise, and control of frequency, coupling, and
pump-phase disorder. The present construction therefore supports
an effective-model emulation route, not a complete hardware
prediction or a proof of equilibrium order in the driven system.

\end{document}